\documentclass[
preprint,
 amsmath,amssymb,
 aps,
]{revtex4-1}
\usepackage{eqnarray,amsmath}
\usepackage{graphicx}% Include figure files
\usepackage{dcolumn}% Align table columns on decimal point
\usepackage{bm}% bold math
\usepackage{hyperref}% add hypertext capabilities
\usepackage{color}
\usepackage{tabularx}
\usepackage{multirow}
\begin{document}

%\preprint{APS/123-QED}

\title{Transitions near the onset of stationary rotating magnetoconvection: role  of magnetic Prandtl number}% Force line breaks with \\

\author{Snehashish Sarkar, Sutapa Mandal,  and Pinaki Pal}
\affiliation{Department of Mathematics, National Institute of Technology, Durgapur~713209, India}

\date{\today}% It is always \today, today,

\begin{abstract}
We investigate the instabilities and associated bifurcation structure near the onset of rotating magnetoconvection of low Prandtl number fluids by performing three dimensional direct numerical simulations. Previous studies considered zero magnetic Prandtl number ($\mathrm{Pm}$) limit for the investigation of bifurcation structure near the onset of convection. Here we numerically investigate the effect of $\mathrm{Pm}$ on the bifurcation structure. The classical Rayleigh-B\'{e}nard convection setup in the presence of horizontal magnetic field and rotation about the vertical axis are considered for the study. The control parameters, including the Taylor number ($\mathrm{Ta}$),  Chandrasekhar number ($\mathrm{Q}$), reduced Rayleigh number ($\mathrm{r}$), and magnetic Prandtl number ($\mathrm{Pm}$) are varied in the ranges $0 < \mathrm{Ta}\leq 500$,  $0 < \mathrm{Q}\leq 1000$, $0.8\leq \mathrm{r} \leq 2$ and $0 < \mathrm{Pm} < 1$ by considering Prandtl numbers $\mathrm{Pr}= 0.025$ and $0.1$. 
The investigation reveals the presence of supercritical, subcritical  and hybrid transitions to convection. These transitions leads to  infinitesimal and finite amplitude fluid patterns at the onset of convection. The finite amplitude solutions can be both stationary and time dependent. The bifurcation structures associated with these flow patterns at the onset are studied in detail. For very small $\mathrm{Pm}$, the bifurcation structure is found to be qualitatively similar to the ones observed in the $\mathrm{Pm}\rightarrow 0$ limit. However, as $\mathrm{Pm}$ is increased, several new solutions appear at the onset and the resulting bifurcation structures are greatly modified. 
\end{abstract}

\maketitle
%\begin{fmtext}
\section{Introduction}
The study of thermal convection in simultaneous presence of rotation and magnetic field, known as rotating magnetoconvection (RMC), is of great importance due to its relevance in geophysical~\cite{marshall:1999, glatzmaier_nature:1999, van_droon_POF:2000, petrelis:2008} and astrophysical~\cite{knobloch_JFM:1981, wickett:2014, miesch:2000, evonuk:2007, hanasoge_ARFM:2016} problems. However, the inherent complexity of these systems due to several factors including geometry leads to the widespread use of the  Rayleigh-B\'enard convection (RBC) model by the researchers to investigate the different properties of RMC. RBC in its simplest form consists of a layer of fluid kept between two horizontal plates maintained at constant temperatures, the lower plate being hotter  than the upper one. The dynamics of RBC is completely described by two dimensionless parameters, namely, the Rayleigh number ($\mathrm{Ra}$, vigor of buoyancy) and the Prandtl number ($\mathrm{Pr}$, ratio of thermal and viscous diffusion time scales). On the other hand, RMC under the paradigm of RBC, along with $\mathrm{Ra}$ and $\mathrm{Pr}$, also needs the dimensionless Taylor number ($\mathrm{Ta}$, strength of rotation),  Chandrasekhar number ($\mathrm{Q}$, strength of the magnetic field) and magnetic Prandtl number ($\mathrm{Pm}$, ratio of magnetic and viscous diffusion time scales) for its description.

Although RMC under Rayleigh-B\'enard geometry models the astrophysical and geophysical systems more closely compared to pure RBC, literature reports only a few works in this direction~\cite{chandra:book, eltayeb_PRS:1972, soward_JFM:1980, aurnou_JFM:2001, aujogue:2015}. Chandrasekhar~\cite{chandra:book} first performed the linear stability analysis of the basic conduction state of RMC of liquid metals with free-slip velocity boundary conditions in the presence of vertical rotation and magnetic field to determine the critical Rayleigh number ($\mathrm{Ra}_c$) and wave number ($\mathrm{k}_c$) for the onset of both stationary cellular and overstable convection in the limit of $\mathrm{Pm}\rightarrow 0$. The choice of the limit $\mathrm{Pm}\rightarrow 0$ is reasonable for the liquid metals, as the value of $\mathrm{Pm}$ is extremely small ($\sim 10^{-7}$) compared to $\mathrm{Pr}~(\sim 10^{-2})$. Chandrasekhar's linear analysis revealed the inhibitory effects of both magnetic field and rotation on the onset of convection (primary instability) which was experimentally verified by Nakagawa~\cite{nakagawa:1957, nakagawa:1959}. Later, in a detailed linear analysis, Eltayeb~\cite{eltayeb_PRS:1972} considered RMC in the presence of external magnetic field of different orientations and the axis of rotation with four sets of boundary conditions. The theoretical analysis determined variety of scaling laws for the onset of convection in the large $\mathrm{Ta}$  and $\mathrm{Q}$ limits. Subsequently, the theoretical investigations performed by Roberts \& Stewartson~\cite{roberts_JFM:1975}, and  Soward~\cite{soward_JFM:1980} on RMC of electrically conducting inviscid fluids in the presence of horizontal magnetic field and vertical rotation along with free-slip and no-slip boundary conditions revealed the existence of oblique rolls and determined their stability regions. The same model was then investigated using linear theory for large Prandtl-number electrically conducting fluids by Roberts and Jones~\cite{roberts_GAFD:2000,jones_GAFD:2000} to determine the preferred mode of convection at the onset. Depending on the ratio of $\mathrm{Pm}$ and $\mathrm{Pr}$, they obtained different flow patterns including  oblique rolls at the onset, which is qualitatively different than the one obtained for inviscid fluids~\cite{roberts_JFM:1975,soward_JFM:1980}.  Thereafter, Aurnou and Olson~\cite{aurnou_JFM:2001} performed an experimental study in the presence of vertical magnetic field and rotation to investigate the heat transfer properties of liquid gallium along with the onset of convection. An interesting theoretical analysis of the similar RMC system was done by Zhang et al.~\cite{zhang:2004} by solving a nonlinear eigenvalue problem and  investigated the effect of electrically
conducting walls on magnetoconvection in a rotating layer of liquid gallium.  Some of the results of the study related to the onset of convection show qualitative match with the Aurnou and Olson's experiment. 

Numerical simulations of a cuboidal cell of liquid metals for the study of flow patterns and heat transfer in the presence of vertical and horizontal magnetic fields as well as vertical rotation have been performed by Varshney and Baig~\cite{Varshney:2008,Baig:2008}. Interesting effects of magnetic field and rotation on the flow patters and heat transfer have been reported. More recently, in a couple of theoretical analysis, preferred mode of convection at the primary instability have been studied for various boundary conditions~\cite{Podvigina:2010,Eltayeb:2013}. Recently, Pal and co-workers investigated the detailed transitions near the onset of rotating magnetoconvection in low Prandtl number electrically conducting fluids by considering the $\mathrm{Pm}\rightarrow 0$ limit both for stationary and oscillatory onset of convection~\cite{nandu_EPL:2015,ghosh_POF:2020,banerjee_PRE:2020,mandal_EPL:2021}. These investigations revealed very rich bifurcation structures involving a large set of local and global bifurcations near the onset of convection. Most of the works discussed above used linear theory and determined the critical Rayleigh number for the onset of convection together with the preferred mode of convection, while some of the works investigated the bifurcation structures and heat transfer properties close to the onset of convection. Almost all the works discussed above, the properties of thermal convection have been studied for very low magnetic Prandtl number fluids by considering the $\mathrm{Pm} \rightarrow 0$ limit. However, very less attention has given by the researchers to investigate the dependence of the bifurcation scenario and related flow structures on magnetic Prandtl number. 

In this paper, we consider RMC in the presence of horizontal magnetic field and rotation about vertical axis, and investigate the effect of the magnetic Prandtl number on the bifurcation structure and the associated flow patterns near the onset of convection by varying $\mathrm{Pm}$ in the range $0 < \mathrm{Pm}<1$. 

\section{Physical system and onset of convection}\label{sec:system}
The classical plane layer Rayleigh-B\'{e}nard convection (RBC) model is considered for the study. It consists of an incompresible homogeneous Newtonian fluid of thickness $h$, kinematic viscosity $\nu$, thermal diffusivity $\kappa$, coefficient of thermal expansion $\alpha$ and magnetic diffusivity $\lambda$, confined between two infinitely extended perfectly conducting parallel plates. The temperature difference between them is $\Delta T = T_l - T_u > 0$, where $T_l$ and $T_u$ respectively are the temperatures of the lower and upper plates and the uniform gravity $g$ acts vertically downwards. The fluid is subjected to an external uniform horizontal magnetic field $\boldsymbol{B_0}=(0,B_0,0)$. The system is uniformly rotated about the vertical axis with an angular velocity $\boldsymbol{\Omega}$ ($= \Omega {\bf {\hat{e}}}_z$), where ${\bf{\hat{e}}}_z$ is the unit vector in the vertically upward direction. When the temperature difference $\Delta T$ between the horizontal plates is less than a critical value, the motionless conduction state prevails in the system which is described by the following equations:
\begin{eqnarray*}
{\bf u}_{cond} &=& 0, \nonumber\\
T_{cond} &=& \mathrm{T}_l - \Delta T \frac{z}{h},\nonumber \\
\rho_{cond} &=& \rho_0( 1 + \alpha \Delta T \frac{z}{h}), \nonumber\\
\pi_{cond} &=& P -\frac{1}{2}|\boldsymbol{\Omega} \times {\boldsymbol r}|^2 - g\rho_0(z + \frac{\alpha \Delta T z^2}{2h}),
\end{eqnarray*}
where ${\bf u}_{cond}$, $T_{cond}$, $\rho_{cond}$, and $\pi_{cond}$ are the velocity, temperature, density and modified pressure fields at the conduction state. $P$ is the pressure field, $\rho_0$ is the reference density, $\alpha$ is the coefficient of volume expansion, and $\boldsymbol{r}$ is the position vector.
As $\Delta T$ crosses a threshold, convective motion starts and the resulting flow of the system under Boussinesq approximation~\cite{boussinesq:1903,boussinesq_spiegel:1960}  is governed by the following set of dimensionless equations:
%\begin{figure}[h]
%\centering
%\includegraphics[scale = 0.6]{./3D.png}
%\caption{Schematic diagram.}
%\label{schematic}
%\end{figure}
\begin{subequations}\label{eq:A}
\begin{gather}
\label{eq:momentum}
\frac{\partial {\bf{u}}}{\partial t} + ({\bf{u}}\boldsymbol{\cdot}\boldsymbol{\nabla}){\bf{u}} = -\boldsymbol{\nabla}{\pi} + \nabla^2{\bf{u}} + \mathrm{Ra} \theta {\bf{\hat{e}}}_z +\sqrt{\mathrm{Ta}}({\bf u} \times {\bf{\hat{e}}}_z)+ \mathrm{Q}\left[\frac{\partial{\bf{b}}}{\partial y} + \mathrm{Pm} ({\bf{b}}\boldsymbol{\cdot}\boldsymbol{\nabla}){\bf{b}}\right], \\ 
\label{eq:heat} 
\mathrm{Pr}\left[\frac{\partial \theta}{\partial t}+({\bf{u}}\boldsymbol{\cdot \nabla})\theta\right] = u_z+\nabla^2\theta ,\\
\label{eq:induction}
\mathrm{Pm}\left[\frac{\partial {\bf{b}}}{\partial t}+({\bf{u}}\boldsymbol{\cdot \nabla}){\bf{b}} - ({\bf{b}}\boldsymbol{\cdot \nabla}){\bf{u}}\right] = \nabla^2 {\bf{b}} + \frac{\partial {\bf{u}}}{\partial y}, \\ 
\label{eq:div_free}
\boldsymbol{\nabla \cdot} {\bf{u}} = 0,~~~\boldsymbol{\nabla \cdot} {\bf{b}} = 0, 
\end{gather}
\end{subequations}
where ${\bf u}(x,y,z,t) = (u_x, u_y, u_z)$, $\theta(x,y,z,t)$, $\pi(x,y,z,t)$ and ${\bf b}(x,y,z,t) = (b_x, b_y, b_z)$ are the convective velocity,  temperature, pressure and induced magnetic field respectively. The above equations are made dimensionless by using the scales $h$, $h^2/ \nu$, $\Delta T \nu /\kappa$ and $\boldsymbol{B_0}\nu/\lambda$ for length, time, temperature and magnetic field respectively. In the non-dimensionalization process five dimensionless parameters introduced here: the Rayleigh number $\mathrm{Ra}=\frac{g\alpha\Delta T h^3}{\nu \kappa}$, the Taylor number $\mathrm{Ta}=\frac{4\boldsymbol{\Omega}^2h^4}{\nu^2}$, the Chandrasekhar number $\mathrm{Q}=\frac{\boldsymbol{B_0}^2h^2}{\nu \lambda \rho_0}$ ($\rho_0$ is the reference density of the fluid), the Prandtl number $\mathrm{Pr}=\frac{\nu}{\kappa}$ and the magnetic Prandtl number $\mathrm{Pm}=\frac{\nu}{\lambda}$.

The top and bottom plates are assumed to be perfectly thermally conducting and stress-free, leading to the conditions
\begin{eqnarray}
u_z = \frac{\partial u_x}{\partial z}=\frac{\partial u_y}{\partial z}={\theta}=0 ~~~\mathrm{at} ~~~ z = 0,1.\label{bc1}
\end{eqnarray}
The horizontal boundaries are also considered to be perfectly electrically conducting, which gives 
\begin{eqnarray}
b_z = \frac{\partial b_x}{\partial z}=\frac{\partial b_y}{\partial z}=0 ~~~\mathrm{at} ~~~ z = 0, 1.\label{bc2}
\end{eqnarray}
Therefore, equations~(\ref{eq:momentum})-(\ref{eq:div_free}) together with the boundary conditions~(\ref{bc1}) and~(\ref{bc2}) provide the mathematical framework to study the considered rotating hydromagnetic system.

Next we use linear theory as described in~\cite{chandra:book, ghosh_POF:2020} to determine critical Rayleigh number ($\mathrm{Ra}_c$) and wave number ($k_c$) for the onset of convection together with the preferred mode of convection. Thus, following~\cite{ghosh_POF:2020}, we consider the linearized versions of the equations (\ref{eq:momentum}) - (\ref{eq:div_free}) and substitute the independent fields in terms of the normal modes of the form
\begin{equation}
A(z)e^{(i(k_x x+k_y y)+\sigma t)},\label{normal_mode}
\end{equation}
where $A(z)$ is the boundary condition compatible $z$-dependent part, $k_x$ and $k_y$ are the horizontal wave numbers respectively.

Subsequently, using the boundary conditions, we obtain the dispersion relation for determining the onset of convection as

\begin{eqnarray}\label{stability_condition}
& (\pi^2+k^2+\mathrm{Pr}\sigma)\left[(\pi^2+k^2)\left((\pi^2+k^2+\sigma)(\pi^2+k^2+\mathrm{Pm}\sigma)+\mathrm{Q}k_y^2\right)^2  + \mathrm{Ta}\pi^2(\pi^2+k^2+\mathrm{Pr}\sigma)^2\right] \nonumber \\
& = -\mathrm{Ra} k^2(\pi^2+k^2+\mathrm{Pm}\sigma)\left[(\pi^2+k^2+\sigma)(\pi^2+k^2+\mathrm{Pm}\sigma)+\mathrm{Q}k_y^2\right],
\end{eqnarray}
with $k = \sqrt{k_x^2 + k_y^2}$, called the wave number of the perturbation. 
In this paper, we are interested to study stationary convection, where the `{\it Principle of exchange of stabilities}'~\cite{chandra:book} is valid. Thus, we set the temporal growth rate $\sigma = 0$ in the equation (\ref{stability_condition}) and obtain
\begin{equation}\label{rayleigh_number}
\begin{split}
\mathrm{Ra}(\mathrm{Ta},\mathrm{Q}) & = \frac{(\pi ^2+k^2)\left[(\pi^2+k^2)^2+\mathrm{Q}k_y^2 \right]}{k^2} +\frac{\mathrm{Ta} \pi^2 (\pi^2+k^2)^2}{k^2 \left[(\pi^2+k^2)^2+\mathrm{Q} k_y^2\right]}.
\end{split}
\end{equation}
The equation (\ref{rayleigh_number}) is then used to determine the critical Rayleigh number $\mathrm{Ra_c}$ along with critical wave number ${k_c}$ for the onset of convection.  As reported in~\cite{ghosh_POF:2020}, the preferred mode of convection determined from the above linear analysis is two dimensional rolls with wave number $k_c$ and it starts to grow at the onset of convection. The saturation and nonlinear dynamics of these growing two dimensional rolls are investigated by performing three dimensional direct numerical simulations. The details are described in the following sections.

\section{Direct Numerical Simulations details}
Linear theory described in the previous section, only determines the growing two dimensional rolls modes along with the critical wave number ($k_c$). However, to determine the flow patterns, which is purely a nonlinear phenomenon,  we perform direct numerical simulations of the system of equations (\ref{eq:momentum}-\ref{eq:div_free}) together with the boundary conditions (\ref{bc1}-\ref{bc2}) using an open source pseudospectral code {\it Tarang} \cite{mkv:code}. The simulations are performed in a square box of size $2\pi/k_c \times 2\pi/k_c \times 1$.  In the simulation code, independent fields are expanded in terms of orthogonal basis functions, compatible with the boundary conditions. Thus, the component of velocity, magnetic and temperature fields are expanded as
\begin{eqnarray}
(u_z,b_z,\theta) &=& \sum _{l,m,n}(W_{lmn}(t),B_{lmn}(t),\Theta_{lmn}(t))e^{i(lk_x x+mk_y y)}\sin(n\pi z), \\
(u_x,u_y,b_x,b_y) &=& \sum _{l,m,n}(U_{lmn}(t),V_{lmn}(t),Bx_{lmn}(t),By_{lmn}(t))e^{i(lk_x x+mk_y y)}\cos(n\pi z),
\end{eqnarray}
where $U_{lmn},~V_{lmn},~W_{lmn},~B_{lmn},~Bx_{lmn},~By_{lmn},~\Theta_{lmn}$ are the Fourier amplitudes and $l,~m,~n$ can be any positive integers. Most of the simulations are done with $32^3$ spatial grid resolutions considering random initial conditions. However, in some specific cases, $64^3$ grid resolutions are used to check the convergence of the solutions. In the process, it is found that the solutions are well resolved with $32^3$ grid resolutions in the considered region of the parameter space. The time advancement in the simulation is done using fourth-order Runge-Kutta method with Courant-Friedrichs-Lewy (CFL) conditions and  time step $\delta t=10^{-4}$. For the convenience of the simulations, we introduce the parameter $\mathrm{r}$, called  reduced Rayleigh number. It is defined by $\mathrm{r} =\mathrm{Ra}/\mathrm{Ra_c}$. We vary $\mathrm{r}$ in small steps for a fixed value of $\mathrm{Q}$, $\mathrm{Ta}$, $\mathrm{Pr}$, $\mathrm{Pm}$ and final values of all the fields are used as the initial conditions for the next value of $r$. The simulations are performed in the region of the parameter space specified by  $0 < \mathrm{Ta}\leq500$, $0< \mathrm{Q}\leq 1000$, $0 < \mathrm{Pm} < 1$, and $ 0.8 < \mathrm{r}\leq 2$ for two values of $\mathrm{Pr}$, namely, $0.025$ and $0.1$. At the outset, we use the critical wave number ($k_c$) obtained from the linear theory and determine the critical Rayleigh number ($\mathrm{Ra}_c$) from the DNS for the validation of the code. A comparison of the critical Rayleigh numbers obtained from the DNS and linear theory are shown in the Table~\ref{Tab1}. From the table,  it is seen that onset of convection determined from the DNS and the linear theory matches quite satisfactorily. 
\begin{table}[h]
\caption{Critical Rayleigh number ($\mathrm{Ra_c}$) and wave number ($k_c$) obtained from the linear theory (LT) along with a comparison of $\mathrm{Ra_c}$  obtained from the DNS for $\mathrm{Q} = 100$ and different $\mathrm{Ta}$.}
\begin{tabularx}{1\textwidth} { 
   >{\centering\arraybackslash}X 
   >{\centering\arraybackslash}X 
   >{\centering\arraybackslash}X
   >{\centering\arraybackslash}X
   >{\centering\arraybackslash}X
   }
 \hline
 \hline
  $\mathrm{Ta}$ & $k_c$ (LT) & $\mathrm{Ra_c}$ (LT) & $\mathrm{Ra_c}$ (DNS) & Error ($\%$) \\
\hline
1 & 2.226 & 659.51 & 660 & 0.0743   \\
 50  & 2.433 & 748.31 & 740 & 1.1105  \\ 
 100 & 2.595 & 826.29 & 810 & 1.9715 \\
 200 & 2.832 & 960.01  & 927 & 3.4385  \\
 300 & 3.012 & 1075.5 & 1028 & 4.4166  \\
 500 & 3.279 & 1274.6 & 1200 &  5.8528 \\
\hline
\end{tabularx}
\label{Tab1}
\end{table} 
Important to note here that the values of $\mathrm{Ra_c}$ and $k_c$ do not depend on the parameters $\mathrm{Q}, \mathrm{Pr}, \mathrm{Pm}$ in the region of the parameter space considered in the paper. Moving ahead, we now perform extensive DNS of the system in the parameter regime of our interest. The details are discussed subsequently. 

\section{Results and Discussions}
The objective of the present study is to investigate the effect of $\mathrm{Pm}$ on the bifurcation structure near the onset of convection by performing extensive DNS in a region of the parameter space where the {\it `Principle of exchange of stabilities}' is valid and the primary instability is always stationary. The investigation starts with the determination of the nonlinear flow patterns just at the onset of convection. 
\begin{figure}[h]
\centering
\includegraphics[scale = 0.45]{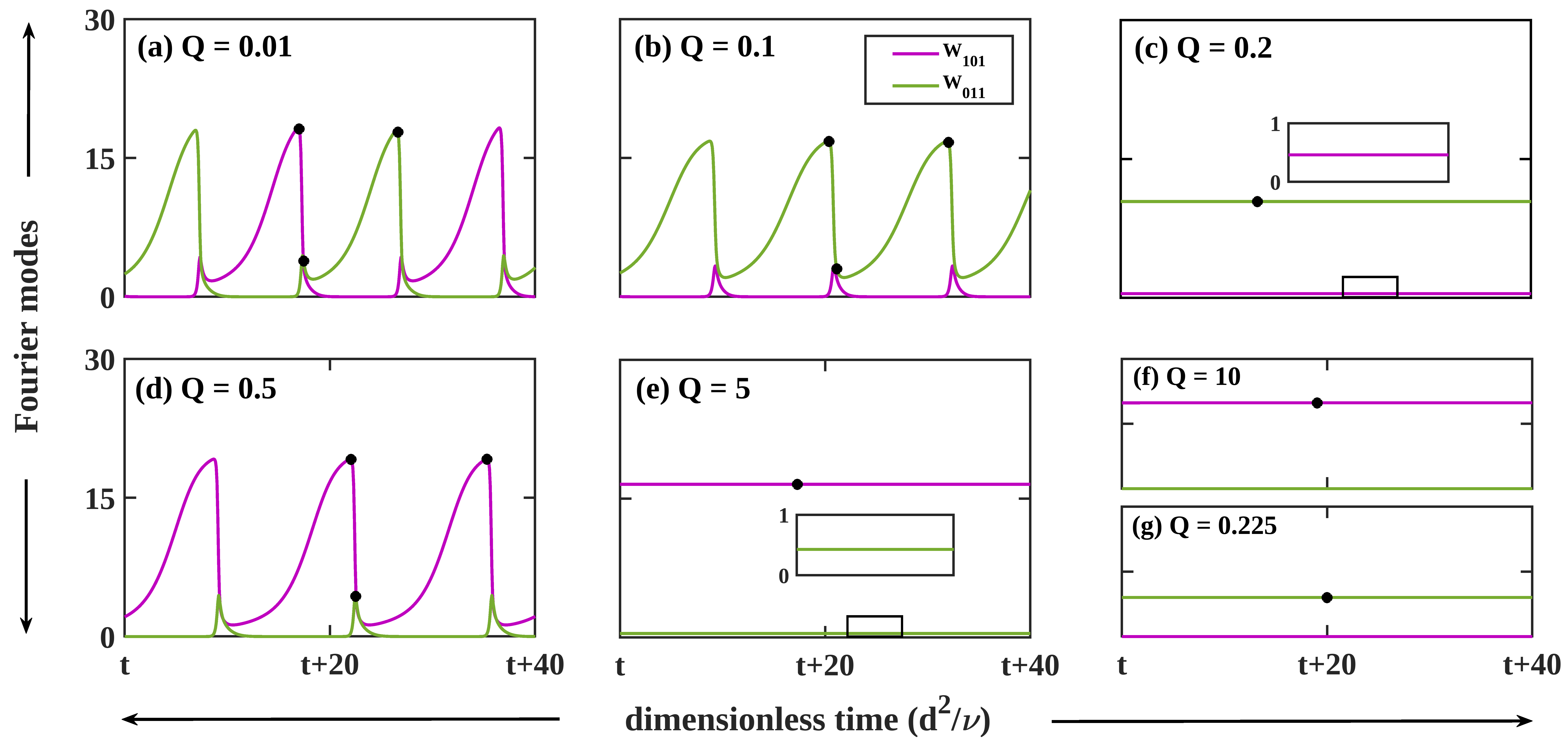}
\caption{Temporal variations of the Fourier modes $W_{101}$ and $W_{011}$ near the onset of convection $(r=1.001)$ for $\mathrm{Pr}=0.025$, $\mathrm{Pm}=10^{-4}$, $\mathrm{Ta}=10$, and different values of $\mathrm{Q}$. (a)-(g) respectively correspond to the OCR-I, OCR-II$'$, CR$'$, OCR-II, CR, SR and SR$'$ solutions respectively. The black dots on the time series mark the time instants at which the isotherms are computed and shown in the figure~\ref{Phase_potrait}.}
\label{Pm0_validation}
\end{figure}

\begin{figure}[h]
\centering
\includegraphics[scale = 0.53]{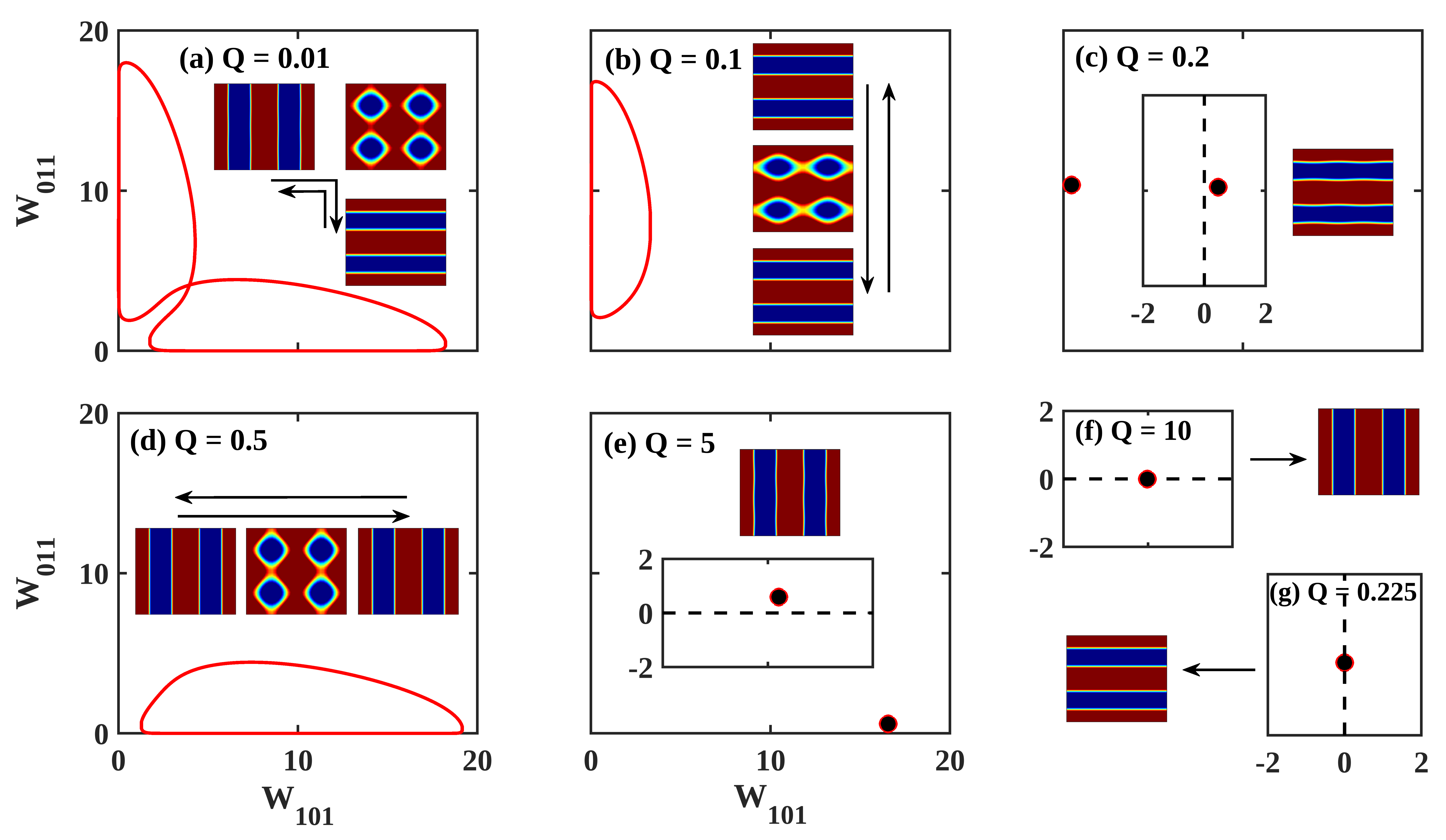}
\caption{(a)–(f) Projections of the phase space trajectories on the $W_{101}-W_{011}$ plane along with the corresponding flow patterns at the onset of convection $(\mathrm{r}=1.001)$ for $\mathrm{Pr}=0.025$, $\mathrm{Pm}=10^{-4}$, $\mathrm{Ta}=10$, and different values of $\mathrm{Q}$. (a), (b) and (d) respectively show the periodic OCR-I, OCR-II and OCR-II$'$ flow patterns, while, (c), (e), (f), and (g) represent the stationary CR, CR$'$, SR and SR$'$ flow patterns respectively, corresponding to the instants marked with the black dots in the figure~\ref{Pm0_validation}.}
\label{Phase_potrait}
\end{figure}

\begin{table}
\caption{Various flow patterns obtained at the onset of convection ($\mathrm{r}=1.001$) from the DNS for $\mathrm{Pr} = 0.025$, $\mathrm{Pm} = 10^{-4}$ and different values of $\mathrm{Ta}$ as a function of $\mathrm{Q}$.}
%Upper, middle and lower blocks are for $\mathrm{Pm} = 10^{-4}$, $10^{-3}$ and $10^{-1}$  respectively. }
\begin{tabularx}{1\textwidth} {  
   >{\centering\arraybackslash}X 
   >{\centering\arraybackslash}X 
   >{\centering\arraybackslash}X
   >{\centering\arraybackslash}X
   >{\centering\arraybackslash}X
   }
 \hline
\hline
Flow Patterns & $\mathrm{Q}(\mathrm{Ta}$ = 10) & $\mathrm{Q}(\mathrm{Ta}$ = 30) & $\mathrm{Q}(\mathrm{Ta}$ = 50) &  $\mathrm{Q}(\mathrm{Ta}$ = 100) \\

\hline
\hline
OCR-I      & $0 - 0.07$     & $0 - 0.44$    & $-$        & $-$          \\
OCR-II$'$  &  $0.08 - 0.16$ & $0.45 - 0.55$ & $-$        & $-$          \\
CR$'$      & $0.17 - 0.29$  & $-$           & $-$        & $-$          \\
SR$'$      & $0.21 - 0.23$ & $-$           & $-$        & $-$          \\
OCR-II     & $0.3 - 2.8$    & $0.56 - 5.4$  & $-$        & $-$          \\
CR         & $2.9 - 7.4$    & $5.5 - 7.5$   & $-$        & $-$          \\
PR-I       & $-$            & $-$           & $-$        & $0 - 3$      \\
SR         & $7.5 - 1000$   & $7.6 - 1000$  & $0 - 1000$ & $4 - 1000$   \\
 \hline
 \hline
 \end{tabularx}
\label{table2}
\end{table}

\subsection{Flow patterns at the onset of convection}
To understand the bifurcation structure near the onset of convection, we first numerically determine the flow patterns just at the onset of convection $(\mathrm{r} = 1.001)$ for the parameter regime specified by $0<\mathrm{Pm} \leq 1$, $0 < \mathrm{Ta}\leq 500$, and $\mathrm{Pr} = 0.025.$ The numerical simulations are started in the very low $\mathrm{Pm}$ regime, and continued for relatively larger $\mathrm{Pm}$. 
\subsubsection{Very small magnetic Prandtl number $(\mathrm{Pm} < 10^{-3})$}
At the outset, we consider $\mathrm{Pr} = 0.025$,  $\mathrm{Ta}=10$ and $\mathrm{Pm} = 10^{-4}$, and numerically investigate the flow patterns very close to the onset of convection  $(\mathrm{r} = 1.001)$  with the variation of $\mathrm{Q}$ using random initial conditions. Figure~\ref{Pm0_validation} shows the time evolution of the Fourier modes $W_{101}$ and $W_{011}$ corresponding to different solutions, as obtained at the onset of convection from the DNS with the variation of $\mathrm{Q}$. 
For very weak external magnetic field, asymmetric self-tune oscillatory solutions (OCR-I) are observed (FIG.~\ref{Pm0_validation}(a)), similar to the ones reported in the $\mathrm{Pm}\rightarrow 0$ limit~\cite{ghosh_POF:2020}. The enhancement of the strength of the magnetic field leads to other solutions at the onset, namely, oscillatory cross rolls (OCR-II/OCR-II$^{'}$), stationary cross rolls (CR/CR$^{'}$) and two dimensional straight rolls (SR/SR$^{'}$) oriented along $x$ or $y$ axes  etc. The time series corresponding to these solutions obtained from the DNS are shown at the FIG.~\ref{Pm0_validation}(b)-(g). The projections of the solution trajectories for these seven different solutions on the $W_{101}-W_{011}$ plane along with flow patterns have been shown in the FIG.~\ref{Phase_potrait}.  Note that the flow patterns are the isotherms computed at the mid-plane $\mathrm{z} = 0.5$.

\begin{figure}[h]
\centering
\includegraphics[scale = 0.55]{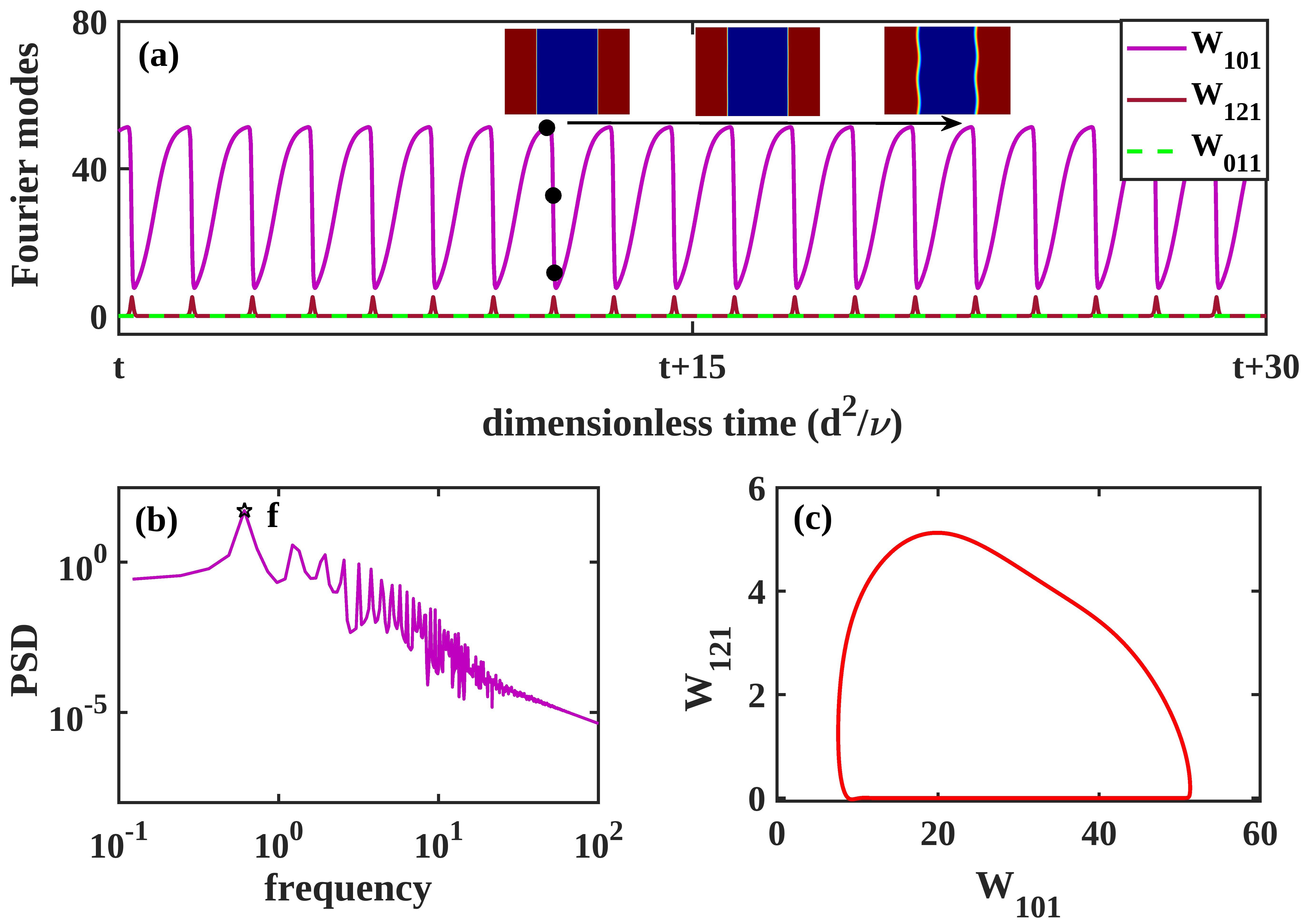}
\caption{(a) Time evolution of the Fourier modes $W_{101}$, $W_{011}$ and $W_{121}$, (b) power spectral density (PSD) and (c) projections of the phase space trajectories on the $W_{101} - W_{121}$ plane for the  periodic rolls solutions (PR-I) obtained at the onset ($\mathrm{r}=1.001$) for $\mathrm{Pr}=0.025$, $\mathrm{Pm} = 10^{-4}$, $\mathrm{Ta}=100$, and $\mathrm{Q}=1$. Note that $W_{011} = 0$ for this type of solution.}
\label{PWR}
\end{figure}

The Chandrasekhar number ranges for the existence of these solutions at the onset of convection for different Taylor numbers are shown in the TABLE~\ref{table2}. In spite of the variation of the flow regimes compared to the $\mathrm{Pm}\rightarrow 0$ results~\cite{ghosh_POF:2020}, a qualitative agreement is observed here.  The table further shows that in the parameter regimes of interest, both stronger magnetic field and rotation suppress the time dependent solutions  and promote stationary two dimensional rolls solutions at the onset. These observations are in line with the results reported in the limit $\mathrm{Pm}\rightarrow 0.$  Therefore, as $\mathrm{Pm}$ approaches $0$, the flow patterns observed at the onset of convection increasingly matches with the ones observed in the $\mathrm{Pm}\rightarrow 0$ limit~\cite{ghosh_POF:2020}. 

Interestingly, for higher rotation rates ($\mathrm{Ta} \geq 100$), periodic rolls of type-I (PR-I) flow patterns, for which $W_{011}=W_{111} =0$ are observed in the presence of weak magnetic field. 
The time series, power spectral density (PSD) and projections on the $W_{101} - W_{121}$ plane for PR-I solution are shown in the figures~\ref{PWR}(a)-(c). The flow patterns for this solution are periodically oscillating rolls, and are shown in ~\ref{PWR}(a) corresponding to the marked instants on the time series. Note that the weak wavyness of the patterns vary periodically with time for this solution.

\begin{table}[h]
\caption{Flow patterns at the onset of convection ($\mathrm{r}=1.001$) for $\mathrm{Pr} = 0.025$, two values of $\mathrm{Pm}$, and four values of $\mathrm{Ta}$ as a function of $\mathrm{Q}$. }
%Upper, middle and lower blocks are for $\mathrm{Pm} = 10^{-4}$, $10^{-3}$ and $10^{-1}$  respectively. }
\begin{tabularx}{1\textwidth} {  
   >{\centering\arraybackslash}X 
   >{\centering\arraybackslash}X 
   >{\centering\arraybackslash}X
   >{\centering\arraybackslash}X
   >{\centering\arraybackslash}X
   >{\centering\arraybackslash}X
   }
 \hline
  \hline
$\mathrm{Pm}$ & Flow Patterns & $\mathrm{Q}(\mathrm{Ta} = 10$) & $\mathrm{Q}(\mathrm{Ta}$ = 30) & $\mathrm{Q}(\mathrm{Ta} = 50$) &  $\mathrm{Q}(\mathrm{Ta} = 100$) \\
\hline
 \hline
          & OCR-I     & $0 - 0.1$    & $0 - 0.9$    & $-$         & $-$ \\
          & OCR-II$'$ & $0.2 - 0.4$  & $1 - 1.3$    & $-$         &    $-$      \\
$10^{-3}$ & OCR-II    & $0.5 - 4.2$  & $1.4 - 8.6$  & $-$         &    $-$      \\
          & PR-I      & $-$          & $-$          & $-$         &    $0 - 30$\\
          & SR        & $4.3 - 1000$ & $8.7 - 1000$ & $0 - 1000$  &  $31 - 1000$\\
 \hline
 \hline
%          &              &        Pm = $10^{-1}$      & &            \\
% \hline
% \hline
          & OCR-I     &  $0 - 0.1$   &  $0 - 0.8$    &    $-$      & $-$ \\
          & OCR-II$'$ & $0.2 - 0.3$  &  $0.9 - 1.5$  &    $-$      & $-$    \\
$10^{-1}$ & OCR-II    & $0.4 - 4.3$  &  $1.6 - 4.6$  &    $-$      & $-$      \\
          & PR-I      & $-$          &  $-$          &    $-$      & $0 - 24$  \\
          & SR        & $4.4 - 1000$ &  $4.7 - 1000$ & $0 - 1000$  & $25 - 1000$\\
\hline
\hline
\end{tabularx}
\label{table3}
\end{table}

Now to investigate the impact of the variation of $\mathrm{Pm}$ we slowly increase the value of it and numerically determine the flow patterns at the onset of convection. The details are discussed subsequently.

\begin{figure}[h]
\centering
\includegraphics[scale = 0.55]{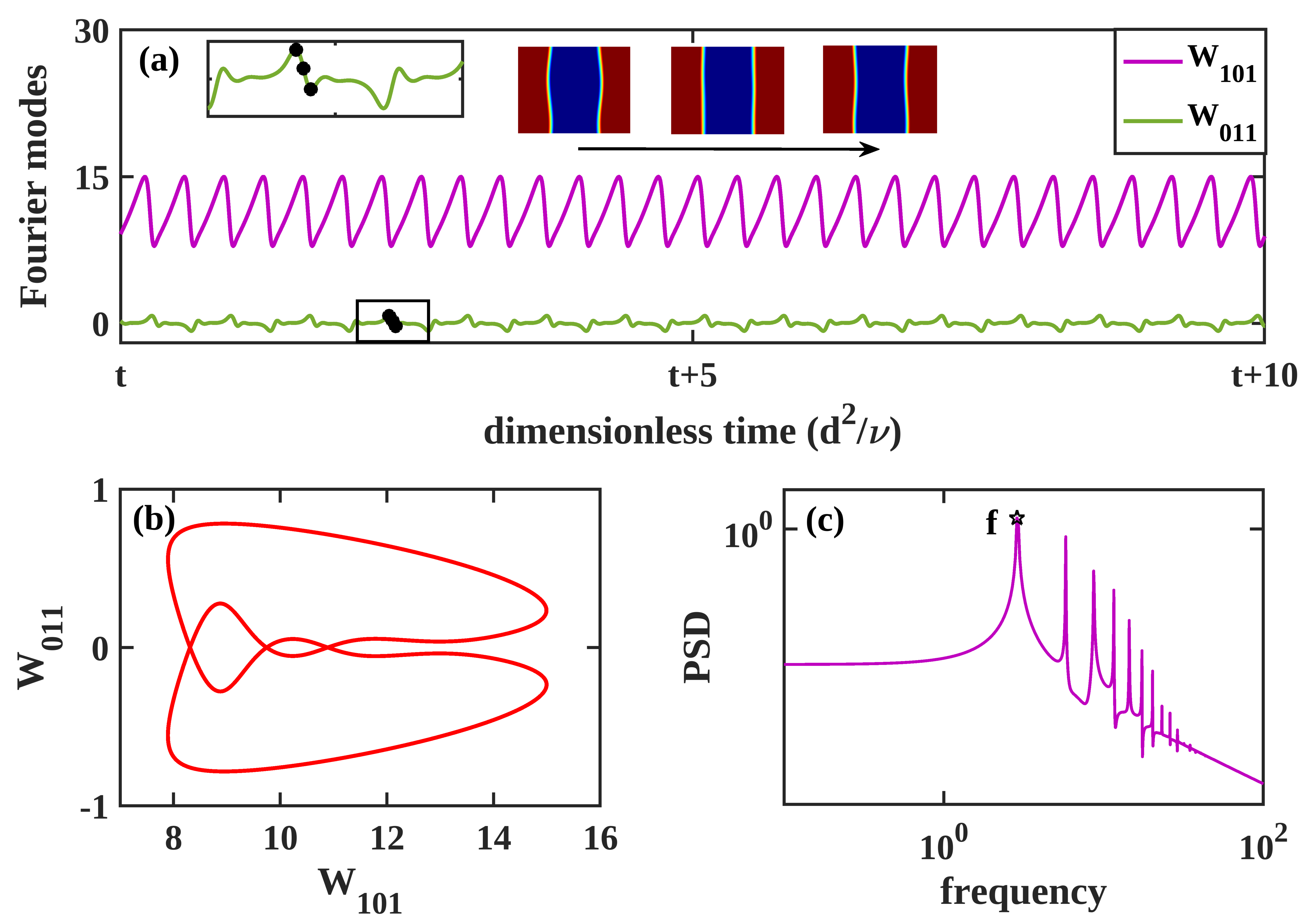}
\caption{(a) Time evolution of the Fourier modes $W_{101}$ and $W_{011}$ along with the flow patterns corresponding to the instants marked by black dots on the time evolution of $W_{011}$, (b) projection on the $W_{101} - W_{011}$ plane, and (c) PSD of $W_{101}$ for the periodic rolls of type II (PR-II) solutions obtained for the parameter values $\mathrm{Pr}=0.025$, $\mathrm{Pm} = 0.5$, $\mathrm{Ta} = 100$, and $\mathrm{Q} = 500$ at the onset of convection  ($\mathrm{r}=1.001$).}
\label{PR}
\end{figure}

\begin{figure}[h]
\centering
\includegraphics[scale = 0.55]{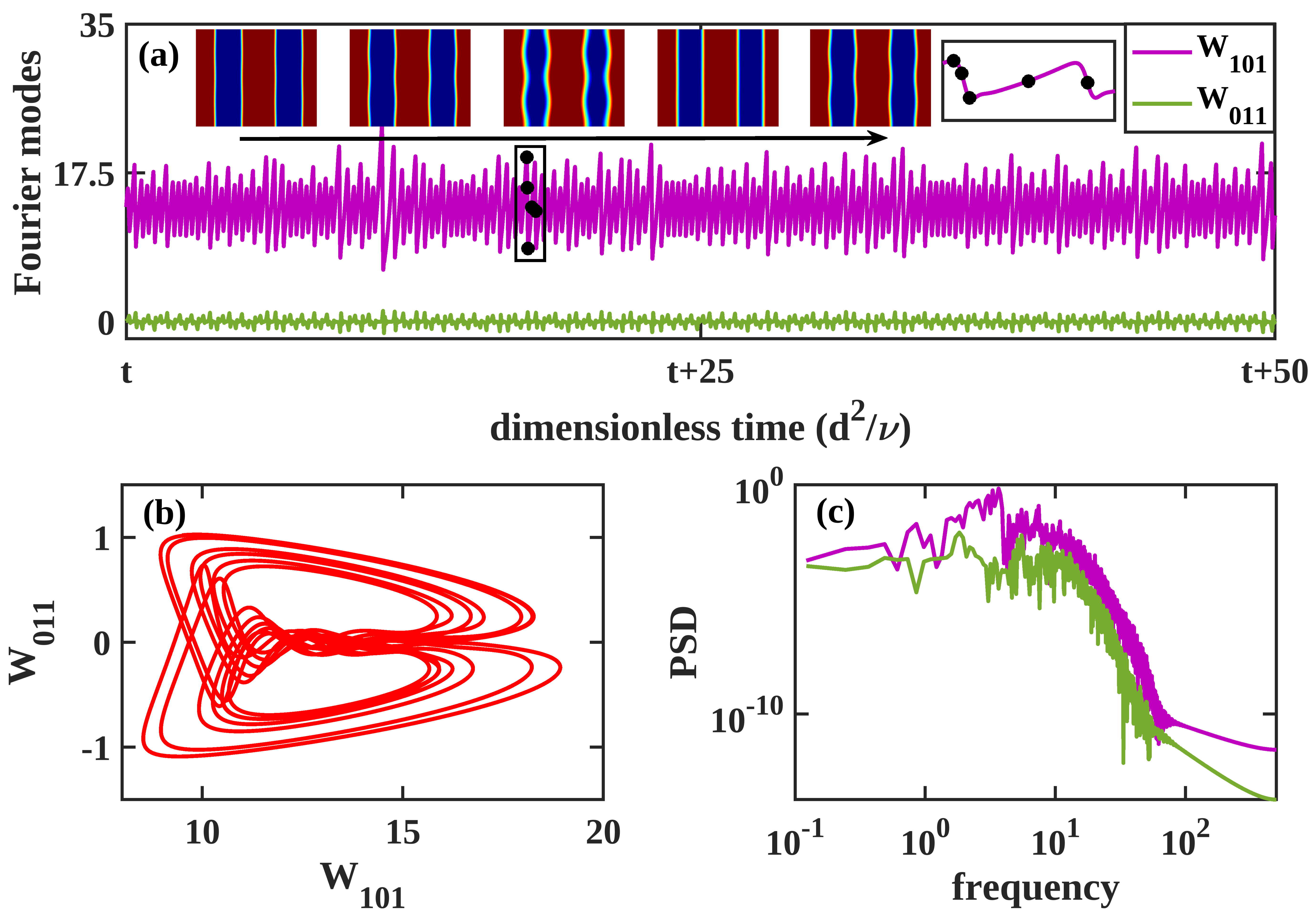}
\caption{(a) Time evolution of the Fourier modes $W_{101}$ and $W_{011}$ along with the flow patterns corresponding to the instants marked by the black dots on the time evolution of $W_{011}$, (b) projection on the $W_{101} - W_{011}$ plane, and  (c) PSD of $W_{101}$ and $W_{011}$ for the chaotic rolls (ChR) solution obtained for the parameter values $\mathrm{Pr}=0.025$, $\mathrm{Pm} = 0.5$, $\mathrm{Ta} = 100$, and $\mathrm{Q} = 800$ at the onset of convection  ($\mathrm{r}=1.001$).} 
\label{CHR}
\end{figure}
 
\subsubsection{Larger magnetic Prandtl number $(10^{-3}\leq\mathrm{Pm}\leq 1)$}
As the value of $\mathrm{Pm}$ increased slowly upto $10^{-1}$, cross rolls (CR and CR$'$) and perpendicular straight rolls (SR$'$) solutions are suppressed. Only five different solutions, namely, OCR-I, OCR-II, OCR-II$'$, PR-I and SR are observed at the onset of convection. Further increase of $\mathrm{Pm}$ brings back the CR solution at the onset for relatively high rotation rate along with the five solutions mentioned before. Interestingly, two new types of solutions, namely, periodic rolls of type II (PR-II) for which $W_{011} \neq 0$ and $W_{111} =0$, and chaotic rolls (ChR) are also observed at the onset. The time series of the modes $W_{101}$ and $W_{011}$ along with the projections on the planes $W_{101}-W_{011}$ plane and flow patterns corresponding to those solutions are shown in the figures~\ref{PR} and \ref{CHR}. It is worth noting that unlike the OCR-II/OCR-II$'$ solutions, for PR-II solutions, $W_{011}$ evolve about zero-mean. 

The details of different flow regimes observed at the onset from the DNS for three magnetic Prandtl numbers are shown in the tables (\ref{table3}-\ref{table4}). The extremely rich flow regimes consisting of ten different flow patterns just at the onset of convection inspire us to determine the type of transitions related to them. 

\begin{table}[h]
\caption{Flow patterns at the onset of convection ($r=1.001$) for $\mathrm{Pr} = 0.025$, $\mathrm{Pm} =5\times 10^{-1}$, and different values of $\mathrm{Ta}$ as a function of $\mathrm{Q}$ . }
\begin{tabularx}{1\textwidth} {  
   >{\centering\arraybackslash}X 
   >{\centering\arraybackslash}X 
   >{\centering\arraybackslash}X
   >{\centering\arraybackslash}X
   >{\centering\arraybackslash}X
   }
 \hline
% \hline
%          &              &    Pm = $5 \times 10^{-1}$          & &            \\
%\hline
\hline
Flow Patterns & $\mathrm{Q}(\mathrm{Ta}$ = 10) & $\mathrm{Q}(\mathrm{Ta}$ = 30) & $\mathrm{Q}(\mathrm{Ta}$ = 50) &  $\mathrm{Q}(\mathrm{Ta}$ = 100) \\
\hline
OCR-I        & $0 - 0.3$     & $0 - 2$   & $-$         & $-$\\
OCR-II$'$    & $0.4 - 0.7$   & $2.1 - 2.5$ & $-$         & $-$\\
OCR-II       & $0.8 - 4.8$   & $2.6 -3.8$& $-$         & $-$\\
PR-I         & $-$           & $-$       & $-$         & $0 - 9$  \\
CR           & $-$           & $-$       & $-$         & $35 - 85$\\
PR-II        & $-$           & $-$       & $-$         & $376 - 624$\\
CHR          & $-$           & $-$       & $-$         & $625 - 925$\\
SR           & $4.9 - 1000$  & $3.9 - 1000$& $0 - 1000$  & $10 - 34$\\
             &               &           &             & $86 - 375$ \\
             &               &           &             & $926 - 1000$ \\
 \hline
 \hline
%          &              &    Pm = $7 \times 10^{-1}$          & &            \\
%\hline
%\hline
% OCR-I       & $0 - 0.6$     &         &             & $-$\\
%OCR-II$'$    & $0.7 - 0.8$   &         &             & $-$\\
%OCR-II       & $0.9 - 1$     &         &             & $55 - 72$\\
%CR           & $-$           &         &             & $30 - 47$\\
%PR           & $-$           &         &             & $541 - 564$\\
%CHR          & $-$           &         &             & $48 - 54$,\\
%             &               &         &             & $565 - 1000$\\
%PWR          & $-$           &         &             & $1 - 15$ \\
%SR           & $10 - 20$     &         &             & $16 - 29$,\\
%             &               &         &             & $73 - 540$\\
%\hline
%\hline
\end{tabularx}
\label{table4}
\end{table}

\begin{figure}[h]
\centering
\includegraphics[scale = 0.55]{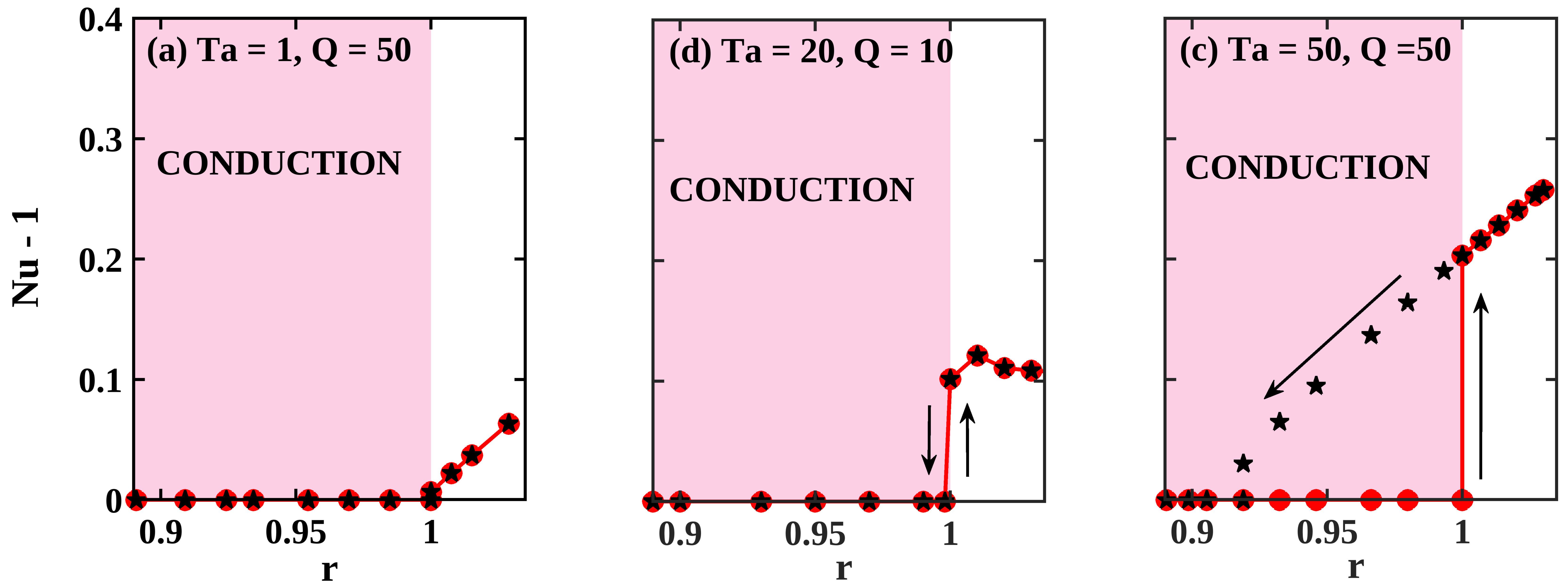}
\caption{Variation of the Nusselt number with $\mathrm{r}$ around the critical point $\mathrm{r} = 1$ for forward (filled red circle) and backward (black star) continuation of solution for different $\mathrm{Ta}$ and $\mathrm{Q}$.  Other parameters are $\mathrm{Pr}=0.025$ and $\mathrm{Pm} =10^{-4}$. (a) shows continuous transition, while, (b) and (c) show discontinuous transitions without and with hysteresis respectively.}
\label{fig:7}
\end{figure}
\begin{figure}
\centering
\includegraphics[scale =0.55]{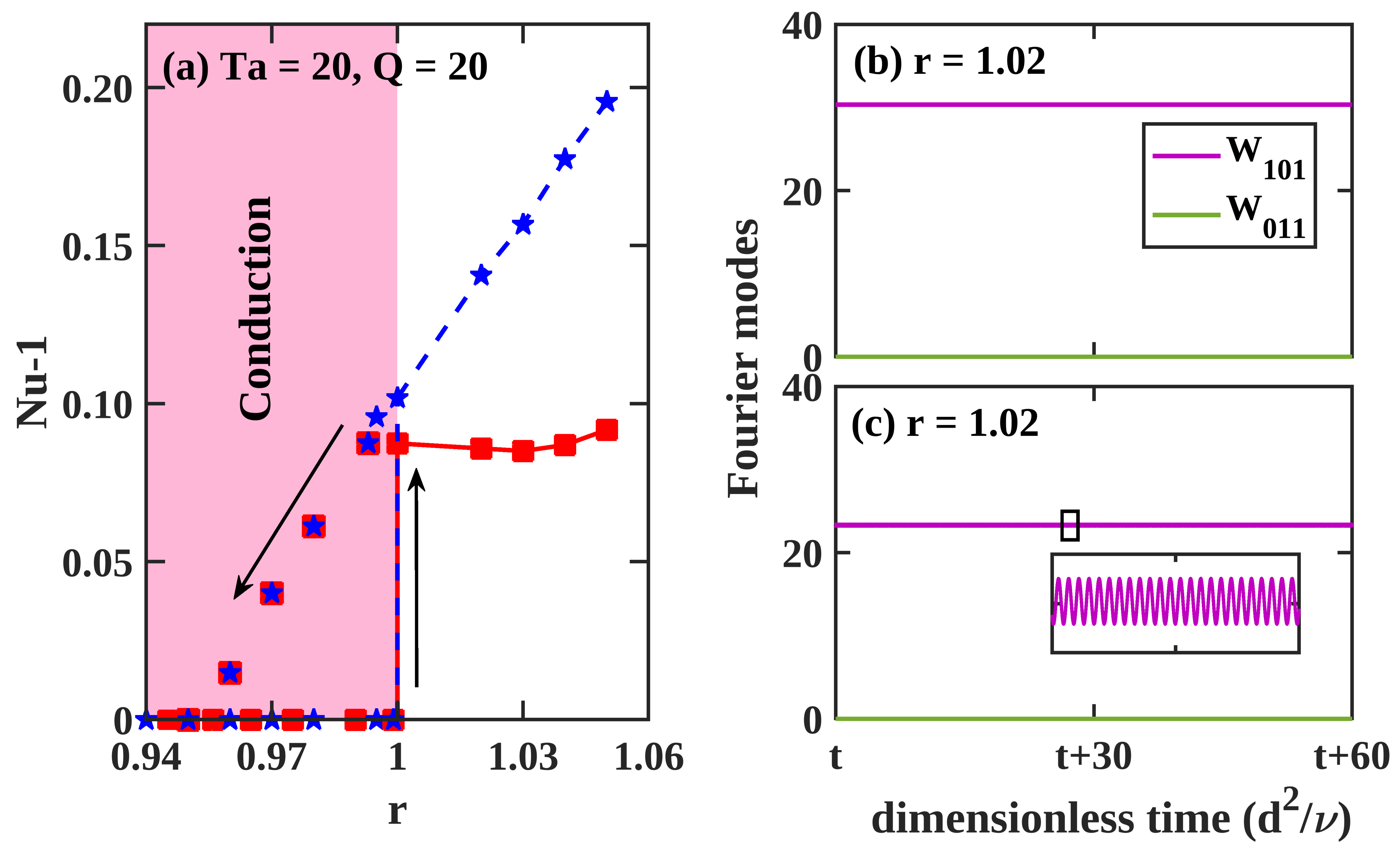}
\caption{(a) Nusselt number ($\mathrm{Nu}$) as a function of $\mathrm{r}$ for two different solutions marked by blue star and filled red squares. (b) and (c) show the time evolution of the Fourier modes $W_{101}$ and $W_{011}$ corresponding to the solutions marked by the blue star and filled red squares in (a).   Forward and backward continuations are indicated by the arrows in (a). The values of the other parameters are $\mathrm{Pm}=10^{-4}$,$\mathrm{Ta}=20$, and $\mathrm{Q}=20$.}
\label{bistable_Nu}
\end{figure} 

\section{Different transitions to convection}
In this section we investigate the type of transitions related to different flow patterns observed at the onset of convection. For that purpose, we compute the Nusselt number ($\mathrm{Nu}$), which is the ratio of total heat flux to conductive heat flux across the fluid layer of the system using the formula
\begin{equation*}
\mathrm{Nu} = 1 + \mathrm{Pr}^2 \langle v_3 \theta \rangle.   
\end{equation*}
Table~\ref{table5} shows the variations of the Nusselt number ($\mathrm{Nu}$) at the onset as a function of $\mathrm{Ta}$ for $\mathrm{Pr} = 0.025$ and different values of $\mathrm{Pm}$. From the tabular data, substantial enhancement of convective heat transfer at the onset of convection for higher rotation rates  is observed. This indicates the possibility of discontinuous or subcritical transition to convection in the considered parameter regime which may lead to finite amplitude solution. To determine the type of transitions, we consider a fixed set of values of $\mathrm{Pr}$, $\mathrm{Pm}$, $\mathrm{Ta}$ and $\mathrm{Q}$, and numerically continue the solutions obtained at the onset with random initial conditions both in forward as well as backward directions in small steps of the reduced Rayleigh number $\mathrm{r}$ and determine the Nusselt numbers. 

\begin{table}[h]
\caption{Nusselt number ($\mathrm{Nu}$) at the onset of convection ($r=1.001$) as a function of $\mathrm{Ta}$ for $\mathrm{Q} = 100$, $\mathrm{Pr} = 0.025$ and different values of $\mathrm{Pm}$. }
\begin{tabularx}{1\textwidth} {  
   >{\centering\arraybackslash}X 
   >{\centering\arraybackslash}X 
   >{\centering\arraybackslash}X
   >{\centering\arraybackslash}X
   }
 \hline
\hline
Ta & Nu (Pm = $10^{-3}$) & Nu (Pm = $10^{-1}$) & Nu (Pm = $5 \times 10^{-1}$)  \\

\hline
\hline
0 & 1.0066 & 1.0066 & 1.0066  \\
20 & 1.0961 & 1.0961 & 1.0961  \\
40 & 1.1711 & 1.1711 & 1.1711  \\
60 & 1.2302 & 1.2302 & 1.2326  \\
80 & 1.2832 & 1.2832 & 1.2996  \\
100 & 1.3252 & 1.3252 & 1.3519  \\
 \hline
 \hline
 \end{tabularx}
\label{table5}
\end{table}

The variation of the Nusselt number with $\mathrm{r}$ across the critical point $\mathrm{r} = 1$ for $\mathrm{Pm} = 10^{-4}$, $\mathrm{Q} = 50$ and for three different $\mathrm{Ta}$ are shown in the figure~\ref{fig:7}.  
The figure~\ref{fig:7}(a) shows that for very low values of $\mathrm{Ta}$, the graph of the Nusselt number is continuous, and the forward and backward continuation data points coincide. Hence the transition to convection is supercritical in nature in this case. As the Taylor number increased in the figures~\ref{fig:7}(b) and (c), discontinuities in the graph of $\mathrm{Nu}$ are clearly observed. In the figure~\ref{fig:7}(b), the forward and backward data points follow the same path but the graph show sudden jump at $\mathrm{r} =1$. Thus, one observes finite amplitude solution at the onset without a hysteresis loop. Such a transition is called hybrid transition~\cite{coutinho:2013, goltsev:2006}. On the other hand, figure~\ref{fig:7}(c) also shows discontinuous transition along with an apparent hysteresis loop. The transition is therefore subcritical in nature. 

Interestingly, for a parameter set we have observed subcritical transition to convection along with bistability at the onset of convection. The variation of the Nusselt number around the critical number $\mathrm{r} = 1$ in this case has been shown in the figure~\ref{bistable_Nu} (a). Two different solutions (marked by red square and blue star in the figure~\ref{bistable_Nu}) are observed for supercritical values of $\mathrm{r}~(> 1)$ depending on the initial conditions and in both the cases the forward and backward continuation data follow different paths showing hysteresis loop. Note that for subcritical values of $\mathrm{r}$ both types of solutions follow the same path. The time series corresponding to two different solutions are shown in the figures~\ref{bistable_Nu}(b) and (c). The solution related to the blue star symbol in the figure~\ref{bistable_Nu}(a) is stationary, while, the one related to filled red square is periodic in nature.  

\begin{figure}[h]
\centering
\includegraphics[scale =0.45]{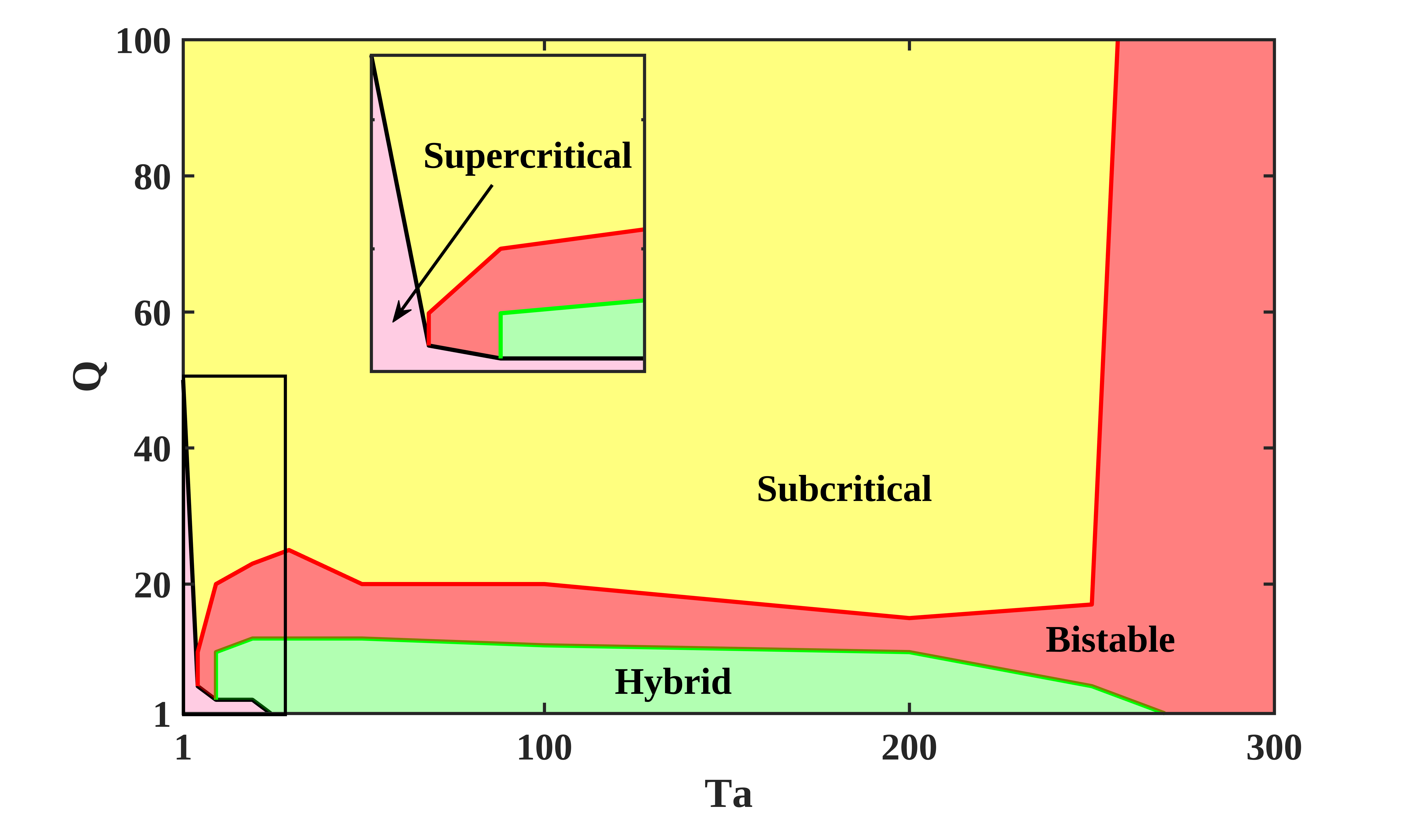}
\caption{Supercritical (pink), subcritical (yellow), hybrid (green) and bistable (red) transition regimes on the $\mathrm{Ta}-\mathrm{Q}$ plane for $\mathrm{Pm}=10^{-4}$ and $\mathrm{Pr}=0.025$. An enlarged view of the marked portion is shown at the inset.}
\label{fig:two_param}
\end{figure}

Inspired by the rich variety of transitions to convection observed in the considered parameter space, we perform extensive DNS in the regime with $\mathrm{Pm} = 10^{-4}$ and $\mathrm{Pr} = 0.025$, and determine the type of transitions to convection on the $\mathrm{Ta} - \mathrm{Q}$ plane. The details of the results obtained from the DNS are presented in the figure~\ref{fig:two_param}. The figure shows the predominance of discontinuous transitions including subcritical (yellow), hybrid (green) and bistable (red) in the parameter regime under consideration leading to the finite amplitude time dependent and stationary solutions at the onset. On the other hand, supercritical transitions leading to small amplitude two dimensional rolls solutions at the onset are observed only for very small $\mathrm{Q}$ and $\mathrm{Ta}$. 

\begin{figure}[h]
\centering
\includegraphics[scale = 0.4]{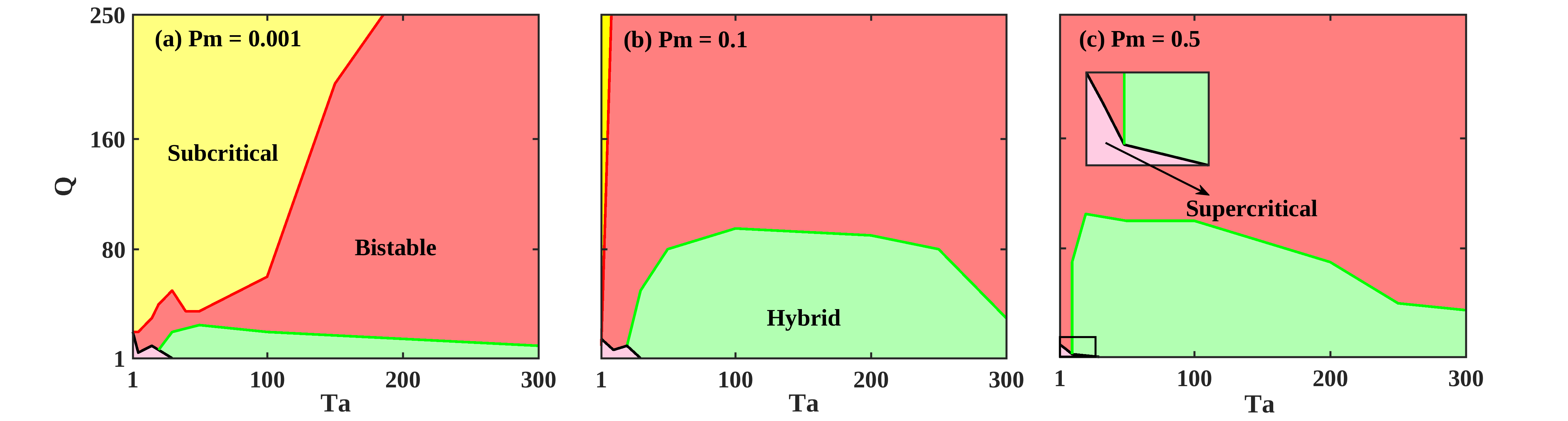}
\caption{Supercritical, subcritical, hybrid and bistable transition regimes  on the $\mathrm{Ta}-\mathrm{Q}$ plane for $\mathrm{Pr}=0.025$ and three $\mathrm{Pm}$. Inset in (c) shows a detailed view of the marked portion.}
\label{fig:two_param_Pm}
\end{figure}
Next we investigate the effect of magnetic Prandtl number on the transitions scenario at the onset. For fixed $\mathrm{Pr} = 0.025$, as the magnetic Prandtl number is increased, the subcritical regime is greatly suppressed, while, the hybrid and bistable regions are enhanced. The details of the effect of $\mathrm{Pm}$ are clearly shown in the figure~\ref{fig:two_param_Pm}.

\begin{figure}[h]
\centering
\includegraphics[scale =0.4]{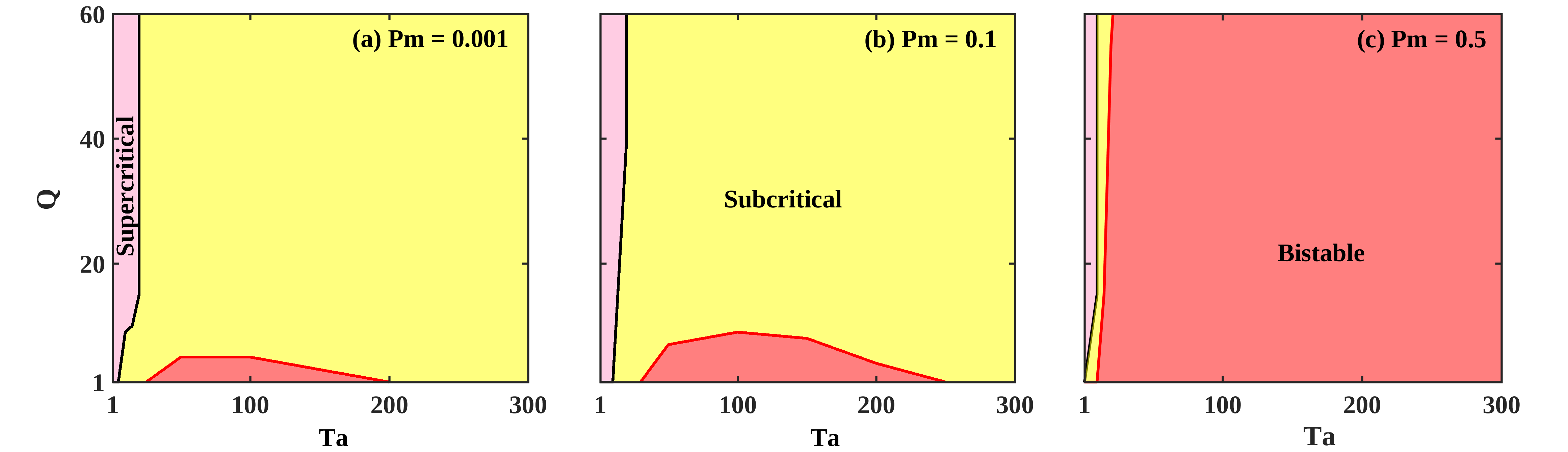}
\caption{Two parameter diagrams showing different transition regimes on the $\mathrm{Ta}-\mathrm{Q}$ plane for $\mathrm{Pr}=0.1$ and three different $\mathrm{Pm}$. The color coding is same as the one used in the FIG.~\ref{fig:two_param_Pm}.}
\label{fig:two_param_pr0p1}
\end{figure}

Further, to investigate the effect of $\mathrm{Pr}$ on the transition scenario, we construct three two parameter diagrams from the simulation data for $\mathrm{Pr} = 0.1$. The diagrams presented in the figure~\ref{fig:two_param_pr0p1} clearly show that the hybrid transition regimes are completely suppressed with the enhancement of the value of $\mathrm{Pr}$.  The supercritical transition regime is slightly increased.  It is also seen that as the value of $\mathrm{Pm}$ increases for $\mathrm{Pr} = 0.1$, the bistable transition regime is greatly enhanced and the subcritical transition regime is greatly suppressed.   

Note that in the two parameter diagrams presented above, we have taken $\mathrm{Q} \geq 1$. For $\mathrm{Q} < 1$ also, we have determined the type of transition and it is observed that the transition is similar to the ones reported  in rotating convection~\cite{mandal:2022} i.e. supercritical and subcritical transitions occur for weak and strong rotation rates respectively. Thus, in this section we discussed the onset of convection through different transition routes including supercritical, subcritical and hybrid. Subsequently we focus on the bifurcation structure associated with different types of solutions of supercritical, subcritical and hybrid origin as the Rayleigh number is increased beyond the onset of convection.

%\begin{figure}[h]
%\centering
%\includegraphics[scale =0.45]{./Q20_Pr0p025_Pm0p001_Supercritical_hybrid_subcritical_New.png}
%\caption{Variation of Nu computed from DNS as a function of r for fixed $\mathrm{Q}=20$, $\mathrm{Pr}=0.025$, $\mathrm{Pm}=0.001$ with three different (a) $\mathrm{Ta}=1$, (b) $\mathrm{Ta}=50$ and (c) $\mathrm{Ta}=60$.}
%\label{fig:Nusselt_number}
%\end{figure}

\section{Flow regimes beyond the onset of convection}
From the tables (\ref{table2} - \ref{table4}), it is apparent that the SR flow patterns are predominant across the regions of the parameter space considered in the paper. Thus, we first investigate the bifurcation structures associated with the SR flow patterns of different origins. 

\subsection{Bifurcation structure associated to SR}

\subsubsection{SR of supercritical origin}
\begin{figure}[h]
\centering
\includegraphics[scale = 0.5]{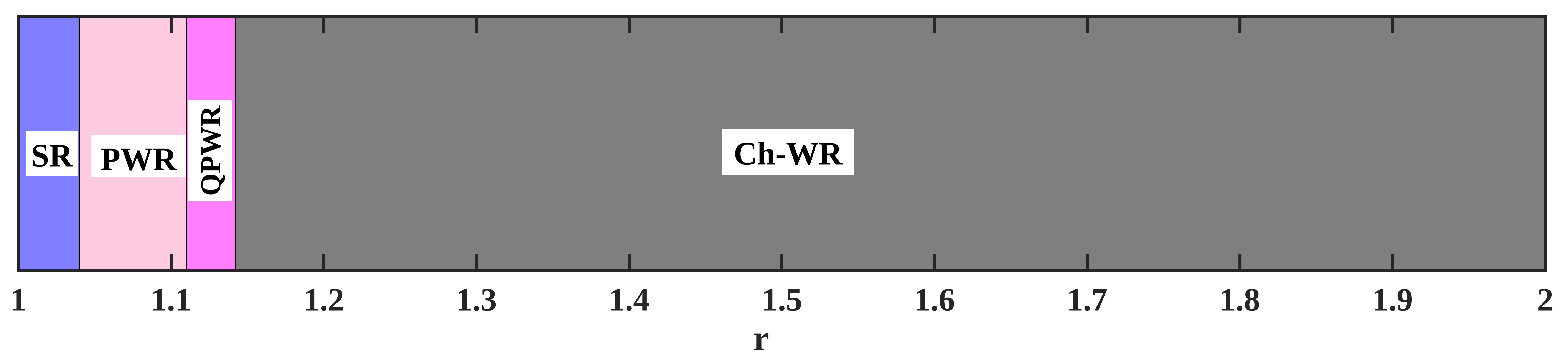}
\caption{Line bifurcation diagram showing different flow regimes associated with the SR flow pattern of supercritical origin using different colors for $\mathrm{Pr} = 0.025,~ \mathrm{Pm} = 10^{-4},~ \mathrm{Q} = 20$ and $\mathrm{Ta} = 1$ with the variation of $\mathrm{r}$.}\label{fig:bar_super_SR}
\end{figure}

The two parameter diagrams presented in the figures~\ref{fig:two_param}, ~\ref{fig:two_param_Pm} and ~\ref{fig:two_param_pr0p1} show the presence of tiny supercritical region where SR flow patterns of supercritical origin are observed. We continue this solution by increasing the value of $\mathrm{r}$ and obtain a quasiperiodic route to chaos. Figure~\ref{fig:bar_super_SR} shows the bifurcation structure associated with the SR flow patterns of supercritical origin for $\mathrm{Pr} = 0.025$, $\mathrm{Ta} = 1$ and $\mathrm{Pm} = 10^{-4}$. As $\mathrm{r}$ is increased, the SR flow patterns become unstable via supercritical Hopf bifurcation at $\mathrm{r} = 1.04$ and periodic wavy rolls (PWR) flow patterns appear in the system. Further increase of $\mathrm{r}$ leads to quasiperiodic wavy rolls through a Neimark-Sacker bifurcation at $\mathrm{r} = 1.11$. The quasiperiodic flow subsequently become chaotic at $\mathrm{r} = 1.142$ with the increment of $\mathrm{r}$. As a result, chaotic wavy rolls (Ch-WR) flow patterns appear and continue to exist till $r = 2.$ For other values of $\mathrm{Pm}$, considered in this paper, the bifurcation scenario associated with the SR flow patterns of supercritical origin is similar. 
\begin{figure}[h]
\centering
\includegraphics[scale = 0.4]{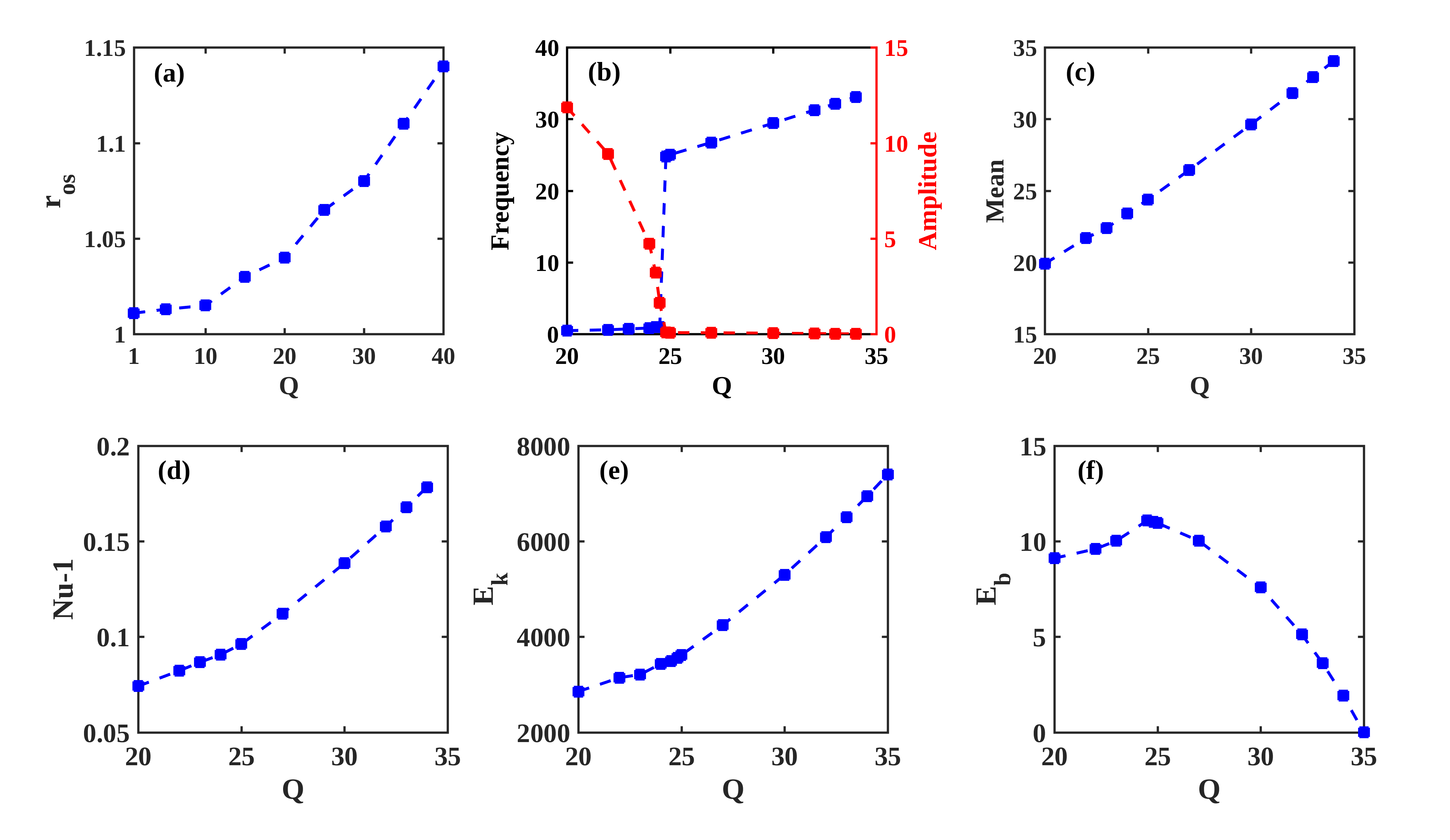}
\caption{(a) The onset of oscillatory instability (OI) of supercritical SR solution as a function of $\mathrm{Q}$ for $\mathrm{Pr}=0.025,~ \mathrm{Pm}=10^{-4}$, and ~ $\mathrm{Ta}=1$. (b) - (f) respectively display the variations of the frequency and amplitude, mean, Nusselt number,  kinetic energy, and  magnetic energy  of the periodic wavy rolls solution obtained at $\mathrm{r} =1.1$ with $\mathrm{Q}$ for the same other parameter values mentioned in (a).}\label{fig:super_Freq_Amp}
\end{figure}

Next we determine the onset of PWR as a function of $\mathrm{Q}$ for $\mathrm{Pr} = 0.025,~ \mathrm{Pm} = 10^{-4}$ and $\mathrm{Ta} = 1$. The critical reduced Rayleigh number $\mathrm{r}_{os}$ for the onset of PWR increase monotonically with $\mathrm{Q}$ (see figure~\ref{fig:super_Freq_Amp}(a)).  Note that $\mathrm{Q} = 1$, PWR solution is observed for $\mathrm{r} = 1.1$. Now keeping  $\mathrm{r} = 1.1$, as $\mathrm{Q}$ is increased, the oscillatory solution is suppressed. The variation of amplitude and  frequency of the PWR with $\mathrm{Q}$ are shown in the figure~\ref{fig:super_Freq_Amp}(b). It is seen that the frequency continuously increases, while, the amplitude of the PWR solutions monotonically increases with $\mathrm{Q}$ before complete suppression for $\mathrm{Q} = 35$. The observation is qualitatively similar to the one observed in the experimental study~\cite{fauve_JPL:1981} in the absence of rotation. The magnetic field suppresses the three dimensional nature of the flow and make it two dimensional. The mean of the oscillatory solution monotonically increases with $\mathrm{Q}$ (figure~\ref{fig:super_Freq_Amp}(c)). We now compute the global quantities like Nusselt number $\mathrm{Nu}$ along with the kinetic energy $\mathrm{E}_k = \frac{1}{2}\langle u_x^2 + u_y^2 + u_z^2\rangle$ and magnetic energy $\mathrm{E}_b = \frac{1}{2}\langle b_x^2 + b_y^2 + b_z^2\rangle$ to see the effect of magnetic field on them. It is evident from the figures~\ref{fig:super_Freq_Amp}(d) and (e) that the external magnetic field facilitates the heat transfer and enhance the kinetic energy in the system. Interestingly, the magnetic energy after the initial growth, continuously decreases with $\mathrm{Q}.$   

\subsubsection{SR of subcritical origin}
\begin{figure}[h]
\centering
\includegraphics[scale = 0.4]{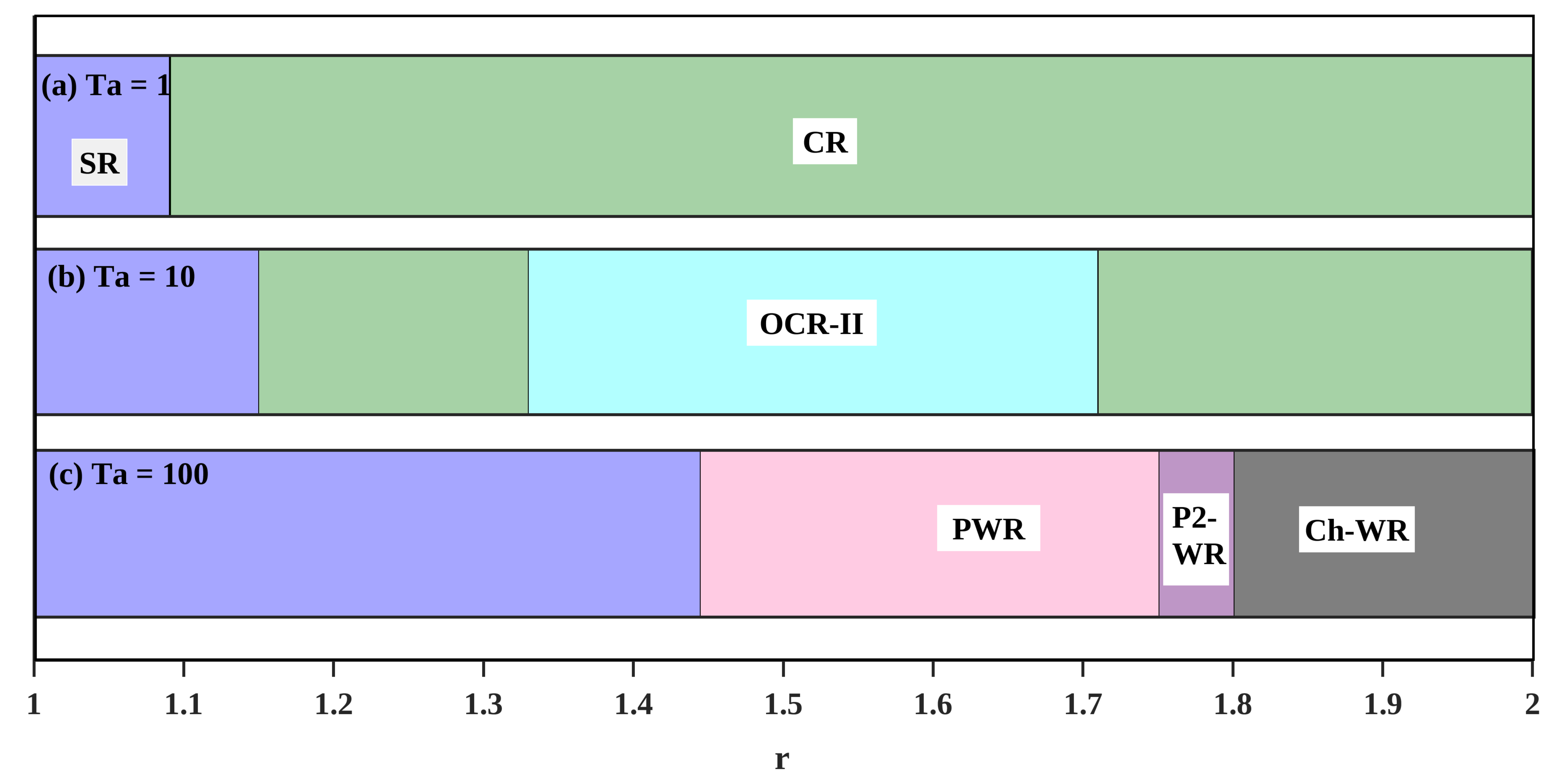}
\caption{Line bifurcation diagrams showing the bifurcation structure associated with the  SR solution of subcritical origin obtained at the onset for $\mathrm{Pr} = 0.025$, $\mathrm{Pm} = 10^{-3}$, $\mathrm{Q} = 100$ and three values of $\mathrm{Ta}$.}\label{fig:bar_sub_SR}
\end{figure}

As shown in the two parameter diagrams presented in the figures~\ref{fig:two_param}, \ref{fig:two_param_Pm}, and ~\ref{fig:two_param_pr0p1}, the straight rolls flow pattern of subcritical origin are observed at the onset of convection in wide regions of the parameter space for all the considered magnetic Prandtl numbers. With fixed $\mathrm{Pr}$, $\mathrm{Pm}$, $\mathrm{Ta}$ and $\mathrm{Q}$, as $\mathrm{r}$ is increased, three distinct scenario are observed for $\mathrm{r} \leq 2$. The scenarios are shown in the bar bifurcation diagrams presented in the figure~\ref{fig:bar_sub_SR}. Figure~\ref{fig:bar_sub_SR}(a) shows that for weak rotation rate, as $\mathrm{r}$ is increased, the straight rolls of subcritical origin bifurcates to a stationary cross rolls (CR) flow patterns at $\mathrm{r} = 1.09$ and continue till $\mathrm{r} = 2.$ With the increment of $\mathrm{Ta}$, the stability region of SR is enhanced and more rich flow regimes are observed as $\mathrm{r}$ is raised. 
\begin{figure}[h]
\centering
\includegraphics[scale = 0.35]{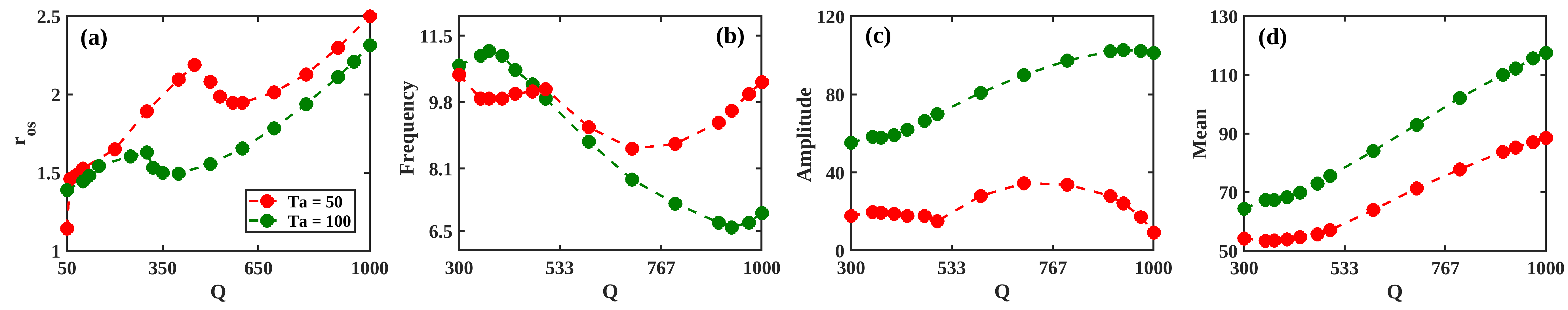}
\caption{(a) The onset of oscillatory instability (OI) of subcritical SR solution as a function of $\mathrm{Q}$ for $\mathrm{Pr}=0.025,~ \mathrm{Pm}=10^{-3}$, and two values of $\mathrm{Ta}$. (b) - (d) respectively display the variations of the frequency, amplitude, and mean of the periodic wavy rolls solution obtained at $\mathrm{r} =2.5$ for $\mathrm{Ta} = 50$ and at $\mathrm{r} = 3.7$ for $\mathrm{Ta} = 100$ with $\mathrm{Q}$ for the same other parameter values mentioned in (a).}
\label{fig:9}
\end{figure}

%\begin{figure}[h]
%\centering
%%\includegraphics[scale = 0.4]{./Ta100_Pm0p001_QvsFrequencyvsAmplitude.png}
%\includegraphics[scale = 0.35]{./Ta100_Pm0p001_QvsFrequency_Amplitude_mean-min.png}
%\caption{Variation of the (a) frequency, (b) amplitude and (c) mean of oscillations with respect to Q, given specific values for $\mathrm{Pr}$ = 0.025, $\mathrm{Pm}$ = 0.001, $\mathrm{Ta}$ = 100 and r = 3.7.  When $\mathrm{Ta}$ = 50, r is set to 2.5, with all other parameters remaining unaltered.}
%\label{fig:10}
%\end{figure}

\begin{figure}[h]
\centering
\includegraphics[scale = 0.35]{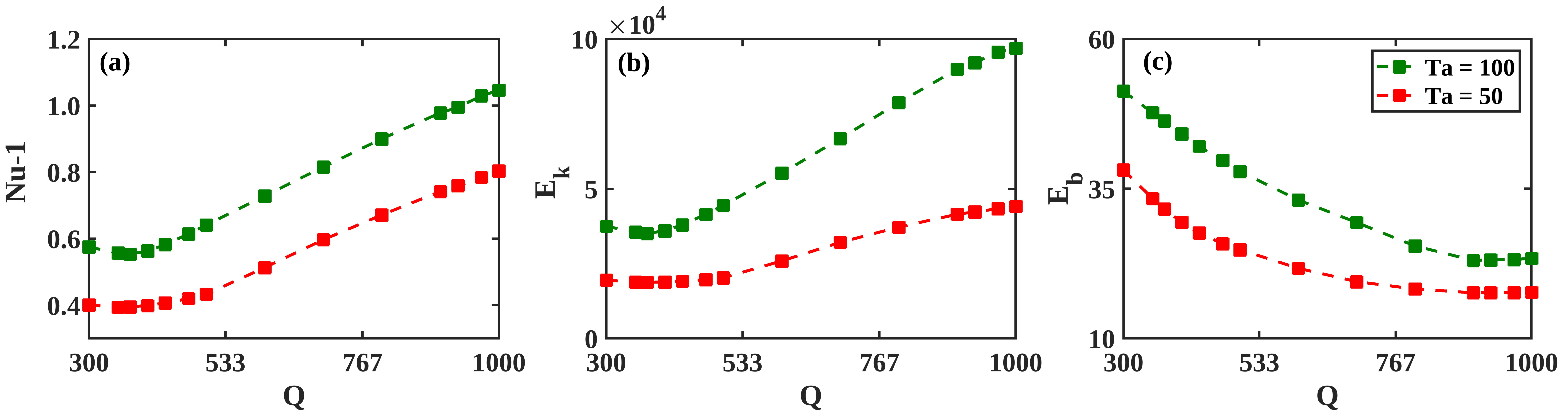}
\caption{Variations of the (a) Nusselt number, (b) kinetic energy, and (c) magnetic energy of the periodic wavy rolls solution with $\mathrm{Q}$ obtained at $\mathrm{r} =2.5$ for $\mathrm{Ta} = 50$ and at $\mathrm{r} = 3.7$ for $\mathrm{Ta} = 100$. Other parameters are $\mathrm{Pr} = 0.025$, and $\mathrm{Pm} = 10^{-3}$.}
\label{fig:11}
\end{figure}
The SR flow patterns become unstable at $\mathrm{r} = 1.15$ and gives rise to the CR flow patterns which once again become unstable via supercritical Hopf bifurcation at $r = 1.33$ leading to periodic oscillatory cross rolls flow patterns of type II (OCR-II) for $\mathrm{Ta} = 10$. A reverse Hopf bifurcation then occurs at $\mathrm{r}  = 1.71$, and stationary CR solutions are generated which continue to exist till $\mathrm{r} = 2.$ The bifurcation sequence is shown in the figure~\ref{fig:bar_sub_SR}(b). Further increase of $\mathrm{Ta}$, leads to a qualitative change in the bifurcation scenario, as shown in the figure~\ref{fig:bar_sub_SR}(c). For $\mathrm{Ta} = 100$, the SR flow regime first become unstable via supercritical Hopf bifurcation at $\mathrm{r} = 1.44$ and gives rise to periodic wavy rolls (PWR) solution with $W_{101}\neq 0$, $W_{011} = 0$ and $W_{111}\neq 0$. As $\mathrm{r}$ is increased, first period-2 wavy rolls (P2-WR) flow patterns and then chaotic wavy rolls (Ch-WR) flow patterns are observed. Thus, the system become chaotic via period doubling route in this case.  

To investigate the transitions related to the SR flow patterns of subcritical origin in more detail, we determine variation of the onset of oscillatory instability of the SR solutions with the magnetic field for two $\mathrm{Ta}$ and show in the figure~\ref{fig:9}(a). The figure shows a non-uniform variation of the onset of oscillatory instability with $\mathrm{Q}$. Initially for weaker magnetic field the onset increases, then a dip is observed followed by a steady increase of $\mathrm{r}_{os}.$ Beyond the onset of oscillatory solution via supercritical Hopf bifurcation of the SR branch we increase $\mathrm{r}$ upto $2.5$ and $3.7$ respectively for $\mathrm{Ta} = 50$ and $100$ for $\mathrm{Q} =300$ where we get PWR solutions. Then we slowly increase the strength of the magnetic field. The solution remains periodic till the highest value of the magnetic field ($\mathrm{Q} = 1000$). The resulting variation of the frequency, amplitude and mean of oscillation with $\mathrm{Q}$ are shown in the figure~\ref{fig:9}(b)-(d). Unlike the supercritical SR case described in the previous subsection, monotonic behavior of the graphs of the  frequency and amplitude are not observed. However, the means of the oscillatory solutions monotonically increase with $\mathrm{Q}.$   Figure~\ref{fig:11} shows the variation of the Nusselt number, kinetic energy and magnetic energy with $\mathrm{Q}$ for two value of $\mathrm{Ta}$. Interesting to note here that for the WR solutions, although the magnetic field helps in the enhancement of heat transfer, and kinetic energy but decreases the magnetic energy of the system. Apart from the SR flow patterns of subcritical origin, one also observes the SR flow patterns related to  the hybrid transition. The forward continuation of this SR flow pattern by increasing the value of $\mathrm{r}$ shows a bifurcation structure similar to the one observed for SR of subcritical origin discussed above.

\subsection{Transitions related to OCR-I, OCR-II and CR}
\begin{figure}[h]
\centering
\includegraphics[scale = 0.45]{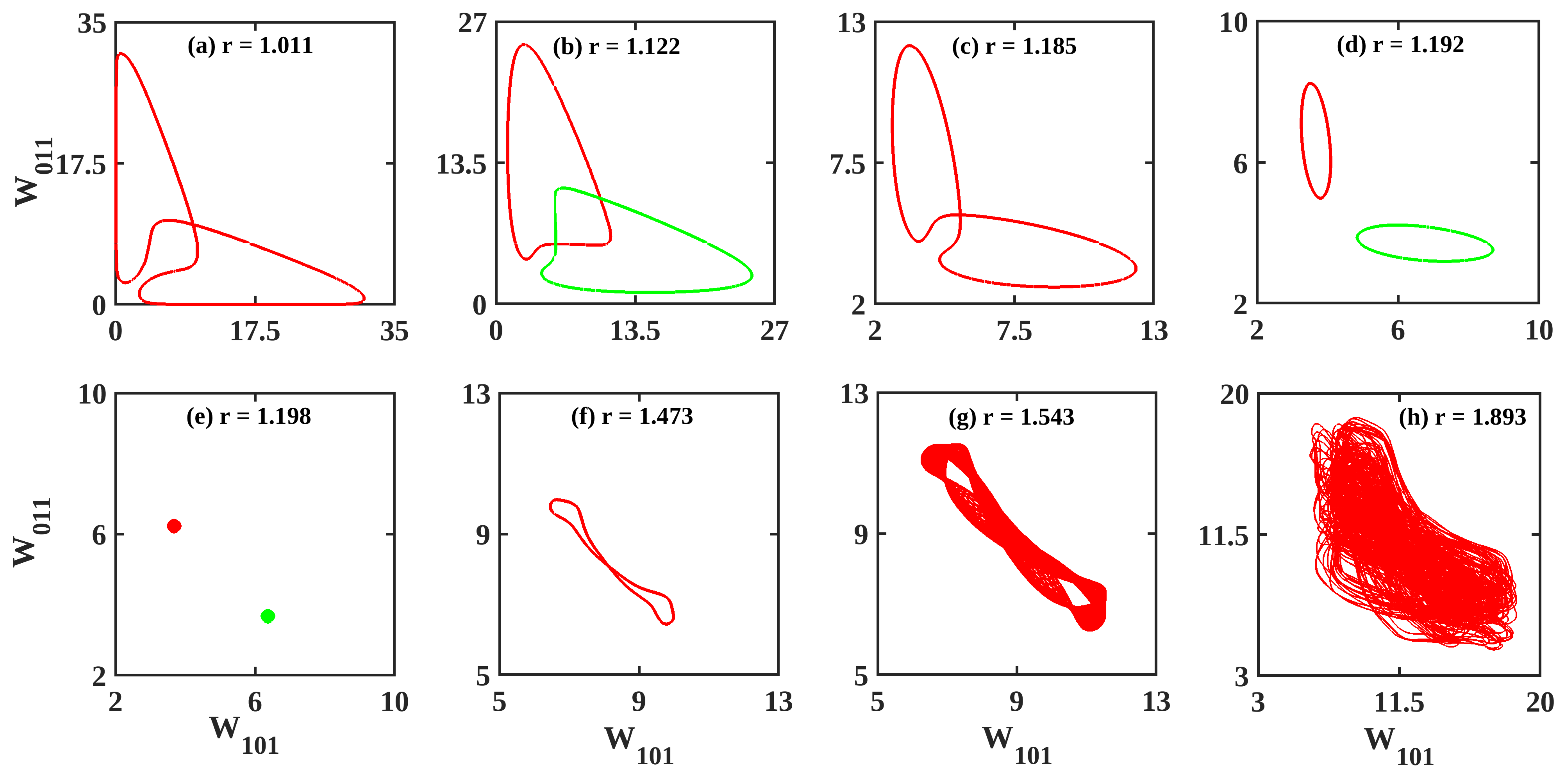}
\caption{Projections of the phase space trajectories on the $W_{101} - W_{011}$ plane with the variation of $\mathrm{r}$ for $\mathrm{Pr} = 0.025$, $\mathrm{Pm} = 0.1$, $\mathrm{Q} = 0.1$ and $\mathrm{Ta} = 30$ showing the gluing and ungluing bifurcations associated with the oscillatory cross rolls solutions. Red and green colors represent two different solutions.}
\label{gluing_ungluing}
\end{figure}

\begin{figure}[h]
\centering
\includegraphics[scale = 0.4]{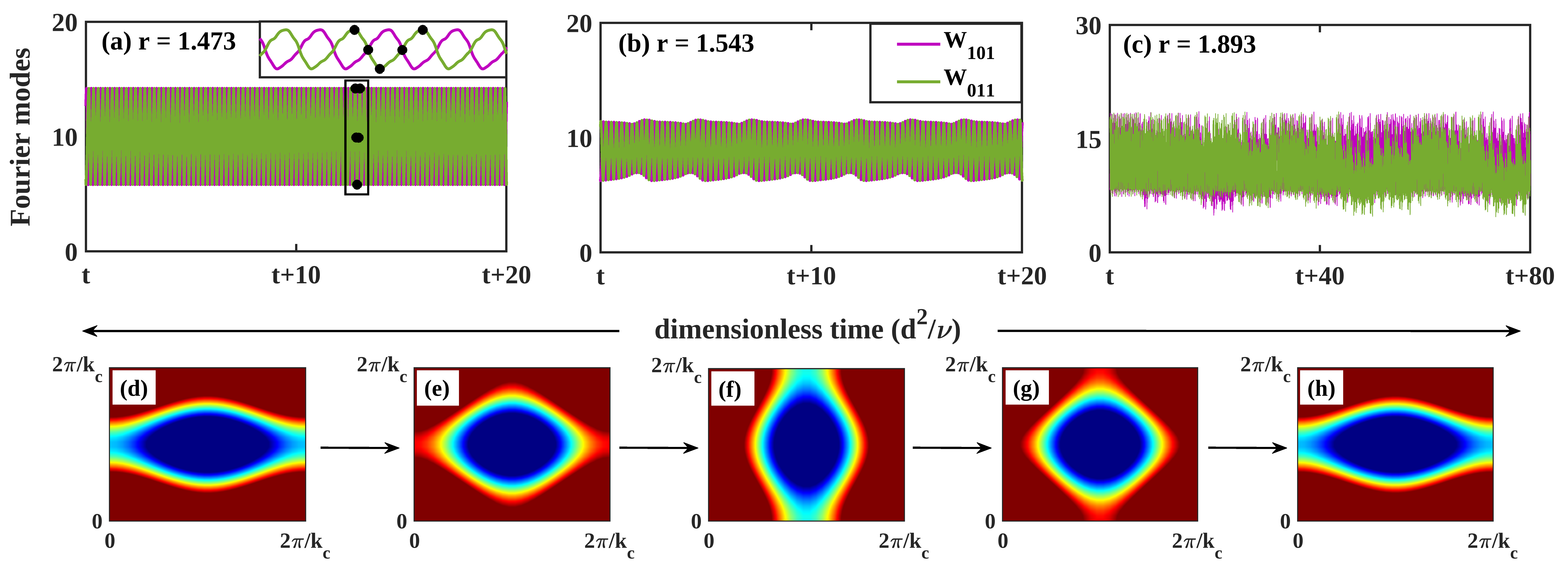}
\caption{(a)-(c) Temporal evolution of the Fourier modes $W_{101}$ and $W_{011}$ for $\mathrm{r} =1.473, 1.543$ and $1.893$ showing the periodic, quasiperiodic and chaotic solutions respectively as observed in the figures~\ref{gluing_ungluing}((f)-(h)). The flow patterns corresponding to the doted instants in (a) are shown in (d)-(h).}
\label{POCR:patterns}
\end{figure}

As evident from the tables~(\ref{table2})-(\ref{table5}) that for very weak magnetic field, OCR-I type of solutions are observed at the onset of convection. To understand the transitions of the OCR-I solutions with the enhancement of $\mathrm{r}$, extensive simulations are performed with $\mathrm{Pr} = 0.025$, $\mathrm{Q} = 0.1$, $\mathrm{Ta} = 30$ and three different values of $\mathrm{Pm}$. In all the cases, OCR-I is observed at the onset of convection and with the increment of $\mathrm{r}$, the system passes through a series of transitions leading to a sequence of flow patterns: OCR-I$\rightarrow$ OCR-II/II$'$ $\rightarrow$ OCR-I $\rightarrow$ OCR-II/II$'$ $\rightarrow$ CR/CR$'$ $\rightarrow$ P-OCR $\rightarrow$ QP-OCR $\rightarrow$ Chaotic.
\begin{figure}[h]
\centering
\includegraphics[scale = 0.4]{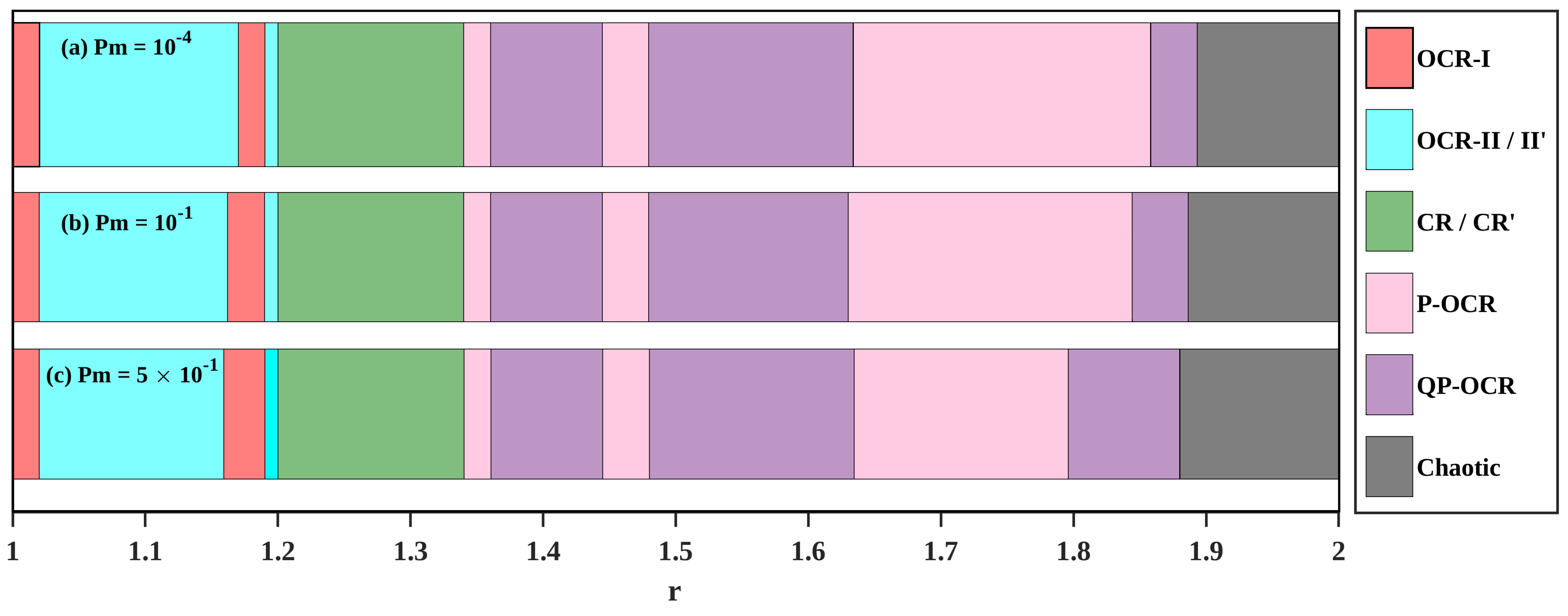}
\caption{Line bifurcation diagrams showing the bifurcation structure associated with the OCR-I solution obtained at the onset of convection for $\mathrm{Pr}=0.025,~ \mathrm{Q}=0.1,~ \mathrm{Ta}=30$ and three different $\mathrm{Pm}$ values.}
\label{OCR-I_tran}
\end{figure}
The details of the transitions of flow patterns are shown in the figure~\ref{OCR-I_tran}. Note that the sequence of flow regimes for three different $\mathrm{Pm}$ is similar. Only the width of the regions of existence of the flow regimes differ with the change of $\mathrm{Pm}$. The projections of the qualitatively different trajectories for $\mathrm{Pm} = 0.1$ are shown in the figure~\ref{gluing_ungluing}. For better understanding of the bifurcation scenario we start bifurcation analysis from the range $1.2 \leq \mathrm{r} \leq 1.34$ where stationary cross rolls either oriented along $x$-axis (CR$'$) or $y$-axis (CR) solutions exist (green regions in the figure~\ref{OCR-I_tran}). As $\mathrm{r}$ is reduced, both the solutions undergo supercritical Hopf bifurcations and oscillatory cross rolls solutions of type II (OCR-II/II$'$) are originated which then merge in a gluing bifurcation leading to OCR-I solution (thin red region in the figure~\ref{OCR-I_tran}). Further reduction of $\mathrm{r}$, brings forth ungluing bifurcation at $\mathrm{r} = 1.16$, and oscillatory cross rolls flow regime of type II starts at $\mathrm{r} = 1.16$ and continue till $\mathrm{r} = 1.02.$    
Finally, once again OCR-I flow regime appears close to the onset of convection through a second gluing bifurcation at $\mathrm{r} = 1.02.$  On the other hand, as $\mathrm{r}$ increased from the CR/CR$'$ regime, the stationary cross rolls solutions become unstable through supercritical Hopf bifurcation and periodic oscillatory cross rolls solutions (P-OCR) appear. The P-OCR solution becomes  quasiperiodic and then chaotic on further increase of $\mathrm{r}$. The projections of the phase space trajectories on the $W_{101} - W_{011}$ plane for qualitatively different flow regimes are shown in the figure~\ref{gluing_ungluing}. The flow patterns corresponding to the OCR-I, OCR-II/II$'$, CR/CR$'$ are already shown in the figure~\ref{Phase_potrait}. The flow patterns corresponding to the P-OCR solution are markedly different than ones shown previously. The time series of the P-OCR solution is shown in the figure~\ref{POCR:patterns}(a). The flow patterns corresponding to the marked instants on the P-OCR solution are shown in the figures~\ref{POCR:patterns}(d)-(h). The time evolution of the quasiperiodic and chaotic solutions arising out of P-OCR solutions on enhancement of $\mathrm{r}$, are shown in the figures~\ref{POCR:patterns}(b)-(c). The flow patterns related to these solutions are qualitatively similar to P-OCR. 
Next we focus on the OCR-II/II$'$ solutions at the onset of convection. The continuation of these flow regimes with the enhancement of $\mathrm{r}$, we observe the sequence: OCR-II/II$'$ $\rightarrow$ OCR-I $\rightarrow$ CR/CR$'$ $\rightarrow$ P-OCR $\rightarrow$ QP-OCR $\rightarrow$ Chaotic of flow regimes. The sequence of flow regimes is similar to the one obtained starting with the OCR-I flow regimes. 
\begin{figure}[h]
\centering
\includegraphics[scale = 0.5]{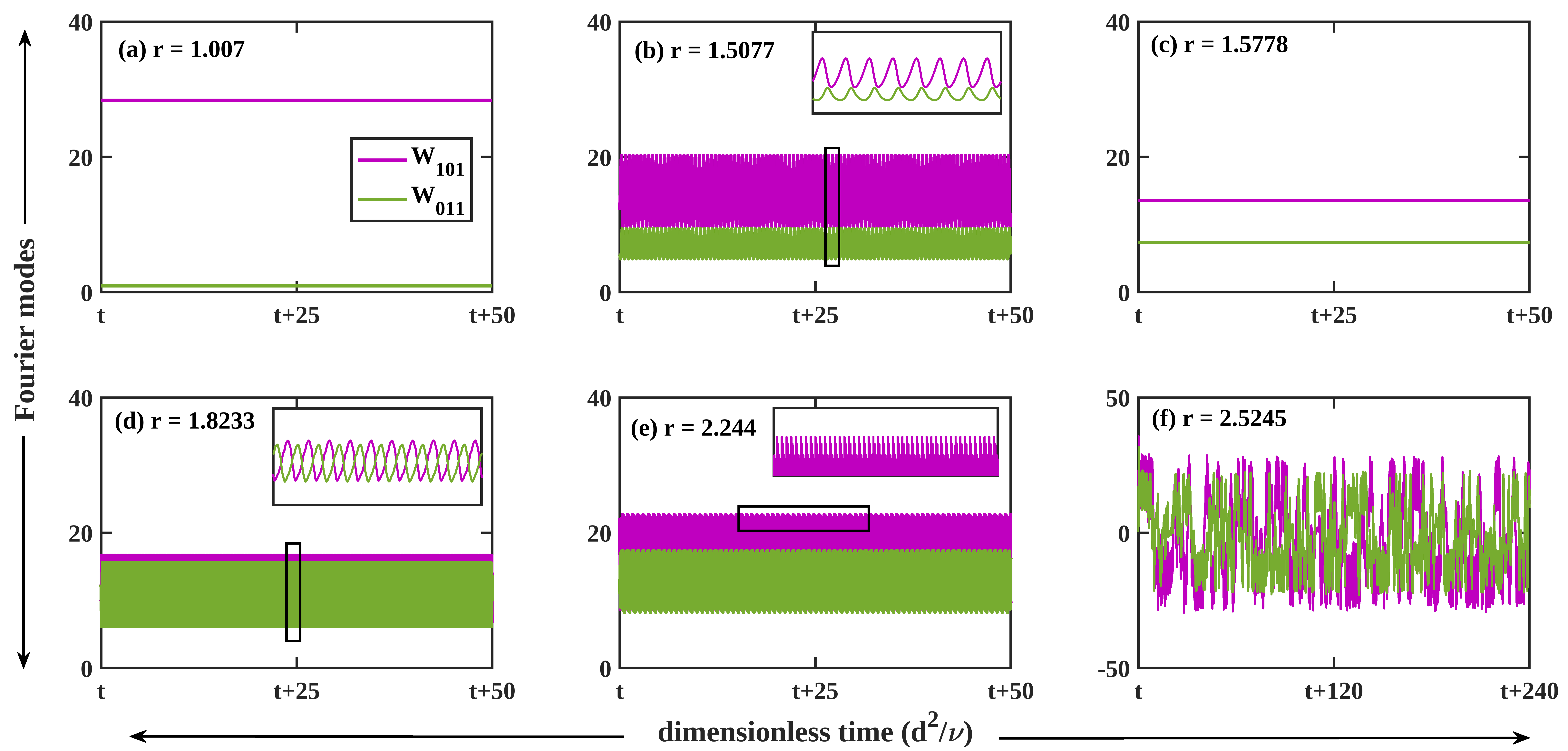}
\caption{Temporal evolution of the Fourier modes $W_{101}$ and $W_{011}$ corresponding to the solutions obtained with the increment of $\mathrm{r}$ for $\mathrm{Pr}=0.025,~\mathrm{Pm}=10^{-4},~\mathrm{Q}=6$, and $\mathrm{Ta}=30$. The enlarged views of the marked regions are shown in the insets.}
\label{CR:Continuation}
\end{figure}
However, the sequence of flow regimes obtained by continuing the CR/CR$'$ flow regimes from the onset of convection by increasing $\mathrm{r}$ shows a minor difference with the ones obtained by continuing the OCR-I and OCR-II type solutions from the onset. In this case, the sequence: CR/CR$'$ $\rightarrow$ OCR-II/II$'$ $\rightarrow$ CR/CR$'$ $\rightarrow$ P-OCR $\rightarrow$ QP-OCR $\rightarrow$ Chaotic of flow patterns are observed. The time evolution of the above mentioned flow patterns are shown in the figure~\ref{CR:Continuation}.

\begin{figure}[h]
\centering
\includegraphics[scale = 0.5]{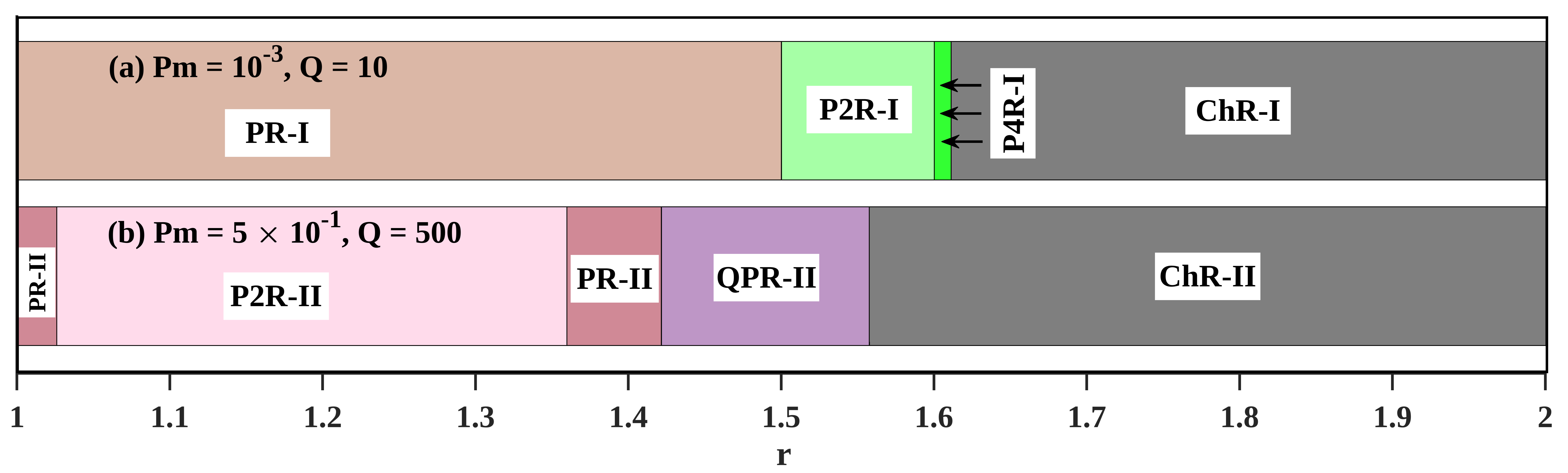}
\caption{Line bifurcation diagrams showing the bifurcation structures associated with the (a) PR-I and (b) PR-II solutions for $\mathrm{Pr}=0.025$ and $\mathrm{Ta}=100$. The other parameters are mentioned inside the figures.}
\label{PR-I:PR-II}
\end{figure}

\subsection{Transitions related to PR-I and PR-II}
As mentioned earlier, for some parameter values the PR-I and PR-II solutions are observed at the onset of convection. Both are of subcritical origin. The continuation of these solutions by increasing the reduced Rayleigh number in small steps, gives rise to two different routes to chaos, namely, period doubling and quasiperiodic. Figure~\ref{PR-I:PR-II} shows the line bifurcation diagrams showing the bifurcation structures associated with the PR-I and PR-II solutions of subcritical origin at the onset of convection. It is observed from the FIG.~\ref{PR-I:PR-II}(a) that with the increase of $\mathrm{r}$, the PR-I solution become unstable at $\mathrm{r} = 1.5$ and a periodic rolls solution of period 2, P2R-I is originated from their. Further increase of $\mathrm{r}$, leads to the appearance of period 4 solution P4R-I at $\mathrm{r} = 1.6$. Finally, it becomes chaotic at $\mathrm{r} = 1.61$ and continue to remain chaotic till $\mathrm{r} \leq 2$. Thus, the route to chaos is period doubling in this case. On the other hand, from the FIG.~\ref{PR-I:PR-II}(b) it is apparent that the continuation of the PR-II solution by increasing $\mathrm{r}$ first leads to the period 2 solution (P2R-II) and it becomes period 1 solution again at $\mathrm{r} = 1.36$. On further increase of $\mathrm{r}$, solution becomes quasiperiodic (QPR-II) at $\mathrm{r} = 1.42$. Subsequently it becomes chaotic at $\mathrm{r} = 1.56$ and remains chaotic till $\mathrm{r} \leq 2$. The route to chaos in this case is quasiperiodic.  

\subsection{Oscillatory instability of rolls}
\begin{figure}[h]
\centering
\includegraphics[scale = 0.45]{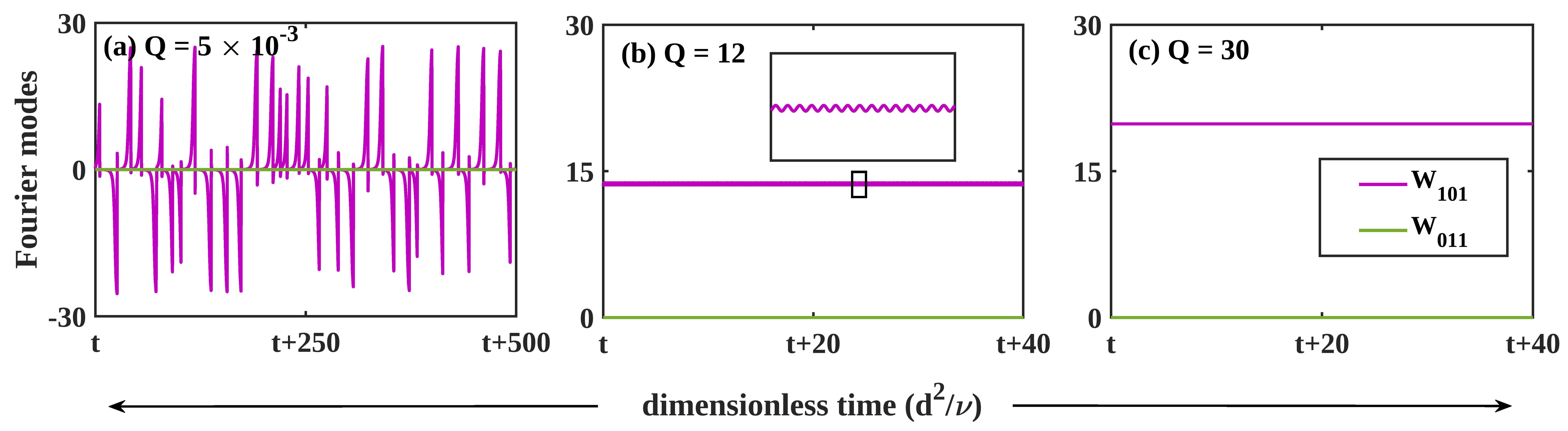}
\caption{Time evolution of the Fourier modes $W_{101}$ and $W_{011}$ for the various flow states that are obtained with the increment of $\mathrm{Q}=30$ for $\mathrm{Pr}=0.025,~\mathrm{Pm}=10^{-4},~\mathrm{Ta}=10$. The inset in the figure (b) shows an enlarged view of the boxed region.}
\label{fig:RollIC_time_series}
\end{figure}

In the preceding subsections we have discussed the bifurcation structures associated with the solutions obtained at the onset of convection using the random initial conditions. However, in this subsection we consider a special set of initial conditions with $W_{101}\neq 0$, $W_{111}\neq 0$ and $W_{011} = 0$, as reported in the paper~\cite{ghosh_POF:2020} and perform extensive DNS. The DNS results show the similar bifurcation structure with the increment of $\mathrm{r}$ as reported in~\cite{ghosh_POF:2020} for very small $\mathrm{Pm}.$ For weak magnetic field, chaotic wavy rolls flow patterns are observed at the onset. As the strength of the magnetic field increased, periodic wavy rolls flow is observed and as the strength of the magnetic field is increased beyond a critical value, time dependence is completely suppressed and steady two dimensional rolls are observed. Figure~\ref{fig:RollIC_time_series} shows the time evolution of different solutions obtained from the DNS at the onset as the value of $\mathrm{Q}$ is increased for $\mathrm{Pr}=0.025,~\mathrm{Pm}=10^{-4},~\mathrm{Ta}=10$. As the value of the reduced Rayleigh number $\mathrm{r}$ increased starting with the chaotic, periodic and stationary solutions at the onset, we observe similar flow regimes as reported in~\cite{ghosh_POF:2020}. Note that as $\mathrm{Pm}$ is increased, we do not observe any qualitative change in the scenario. 
Thus, we focus on the range of $\mathrm{Q}$ where stationary two dimensional flow patterns are observed at the onset. We then increase the value of $\mathrm{r}$ and observe a supercritical Hopf bifurcation leading to oscillatory solution. The threshold value of $\mathrm{r}$ for the onset of oscillatory solution is denoted by $\mathrm{Ra_{o}(\mathrm{Q})}$. The effect of magnetic Prandtl number and rotation rate on the variation of $\mathrm{Ra_{o}(\mathrm{Q})} - \mathrm{Ra}_c$ with $\mathrm{Q}$ is then investigated in detail. Interestingly, the quantity $\mathrm{Ra_{o}(\mathrm{Q})} - \mathrm{Ra}_c$ is found to scale as $\mathrm{Q}^{\gamma}$ for weak rotation rate. 
\begin{figure}[h]
\centering
\includegraphics[scale = 0.45]{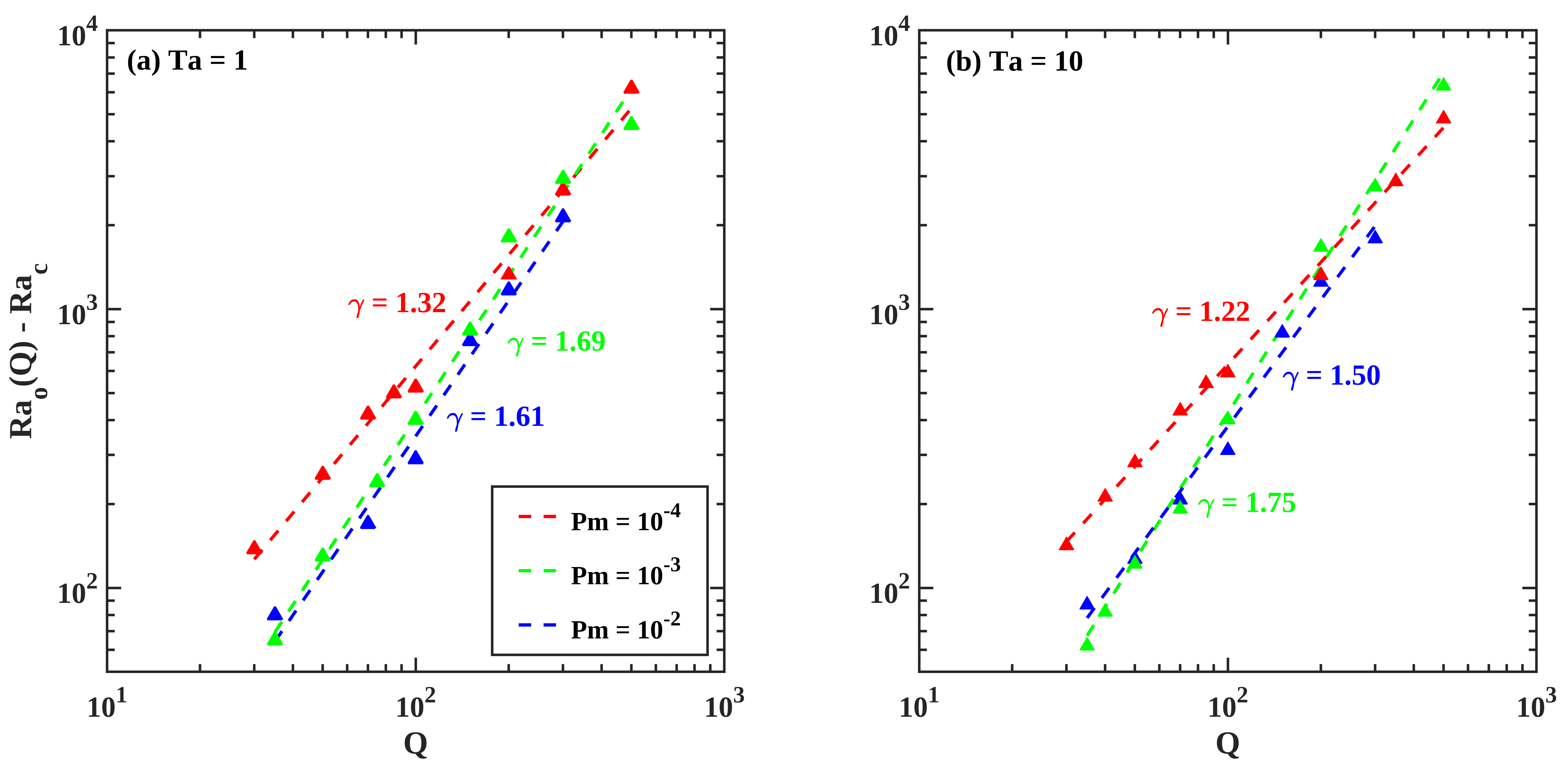}
\caption{The scaling of the onset of oscillatory instability $\mathrm{Ra_{o}(\mathrm{Q})} - \mathrm{Ra}_c$ as a function of $\mathrm{Q}$ for three values of $\mathrm{Pm}$ and two different $\mathrm{Ta}$. The dashed lines represent the corresponding best linear fits.}
\label{fig:scaling}
\end{figure}
Figure~\ref{fig:scaling} clearly shows the said scaling for two $\mathrm{Ta} = 1$ and $10$.  For fixed $\mathrm{Ta}$, the exponent $\gamma$ varies significantly with $\mathrm{Pm}$. Note that scaling law breaks down for higher rotation rate $\mathrm{Ta}$ and $\mathrm{Pm}$. 

\section{Conclusions}
In this paper we have investigated the effect of the magnetic Prandtl number on the bifurcation structure near the onset of rotating convection in low Prandtl number fluids in presence of external magnetic field considering the classical plane layer Rayleigh-B\'enard convection model. A wide region of the parameter space defined by $0 < \mathrm{Ta}\leq 500$,  $0 < \mathrm{Q}\leq 1000$, $0.8\leq \mathrm{r} \leq 2$, $0 < \mathrm{Pm} < 1$ and $\mathrm{Pr} \in \{0.025, 0.1 \}$ have been numerically investigated for the purpose by performing three dimensional direct numerical simulations.  In the region, the `Principle of exchange of stabilities' is valid and stationary cellular pattern of convection is observed at the onset.

The investigation reveals more than ten different types of flow patterns just at the onset of convection and the transitions are found to be continuous and discontinuous. The continuous transition to convection leads to flow patterns of infinitesimal amplitude at the onset, while, the discontinuous transitions lead to finite amplitude flow patterns. The finite amplitude solutions may or may not accompany the phenomenon of hysteresis leading to subcritical and hybrid transitions respectively.  Note that the explored parameter regime is dominated by subcritical and hybrid transitions. 

For very low magnetic Prandtl numbers ($<10^{-3}$), the flow regimes and the associated bifurcation structure is found to be similar to the ones observed in the $\mathrm{Pm}\rightarrow 0$ limit. The bifurcation structure is greatly modified with the enhancement of $\mathrm{Pm}$ in terms of the flow regimes associated with the supercritical, subcritical and hybrid transitions to convection.  For $\mathrm{Pr} = 0.025$, the bistable flow regimes are greatly enhanced with the increment of $\mathrm{Pm}$. The forward continuation of the different solutions observed at the onset of convection by increasing the Rayleigh number in small steps leads to chaotic solutions via either quasiperiodic or period doubling routes. 

Besides, we have also investigated the effect of magnetic Prandtl number on the onset of oscillatory instability of the two dimensional rolls. In a wide range of $\mathrm{Q}$, an well defined scaling law is observed for $\mathrm{Pm} \leq 10^{-2}$ in the presence of weak rotation rate which shows resemblance with the similar scaling laws reported in magnetoconvection system~\cite{fauve_JPL:1981}. Both stronger rotation rate and larger magnetic Prandtl number destroy the scaling law.

\begin{acknowledgements}
SM and SS acknowledge the supports from CSIR India (File No. 09/973(0024)/2019-EMR-I) and UGC India (Award No. 191620126754) respectively. The authors thankfully acknowledge the support of Manojit Ghosh in direct numerical simulations.
\end{acknowledgements}

%\bibliographystyle{unsrt}
%\bibliography{sources}

\end{document}